\documentclass[journal]{IEEEtran}

\normalsize

\ifCLASSINFOpdf

\else

\fi

\usepackage{epsfig,makeidx,color,mathtools,bigints,graphicx,amsbsy,amsmath,amssymb,euscript,verbatim}
\usepackage{url}
\DeclareMathOperator{\argmin}{argmin}

\DeclareMathOperator{\SINR}{SINR}
\DeclareMathOperator{\MSE}{MSE}

\DeclareMathOperator{\vol}{vol}
\DeclareMathOperator{\dg}{dg}
\DeclareMathOperator{\tr}{tr}

\newtheorem{prop}{Proposition}
\newtheorem{theorem}{Theorem}
\newtheorem{lemma}{Lemma}

\begin{document}

\title{Channel Pre-Inversion and max-SINR Vector Perturbation for Large-Scale Broadcast Channels}

\author{David A. Karpuk, \IEEEmembership{Member, IEEE,} 
	and Peter Moss
\thanks{D.\ Karpuk is with the Department of Mathematics and Systems Analysis, Aalto University, Espoo, Finland.  P.\ Moss is formerly of BBC Research and Development, London, United Kingdom, and is currently an independent consultant.  emails: david.karpuk@aalto.fi, pnm30@hotmail.com.}
\thanks{D. Karpuk is supported by Academy of Finland Postdoctoral Researcher grant 268364.}}

\maketitle
\IEEEpeerreviewmaketitle

\begin{abstract}
We study channel pre-inversion and vector perturbation (VP) schemes for large-scale broadcast channels, wherein a transmitter has $M$ transmit antennas and is transmitting to $K$ single-antenna non-cooperating receivers.  We provide results which predict the capacity of MMSE pre-inversion as $K\rightarrow\infty$.  We construct a new VP strategy, \emph{max-$\SINR$ vector perturbation} (MSVP), which maximizes a sharp estimate of the signal-to-interference-plus-noise ratio.  We provide results which predict the performance of MSVP and demonstrate that MSVP outperforms other VP methods.  Lastly, we combine MSVP with the low-complexity Sorted QR Precoding method to show that MSVP has the potential to efficiently deliver data to a very large number of users at close to channel capacity. 


\end{abstract}

\begin{IEEEkeywords}
Channel Pre-inversion, MMSE Inverse, Vector Perturbation, SINR, Random Matrix Theory, Sorted QR Precoding, Broadcast Channels
\end{IEEEkeywords}

%
\IEEEpeerreviewmaketitle

\section{Introduction}

Successful implementation of next-generation (e.g.\ 5G) mobile broadband internet will require the delivery of high-volume and high-fidelity data (e.g.\ streaming video) simultaneously to a large number of users.  The rapid increase of both the number of mobile devices and the volume of data to be delivered is putting heavy demands on broadcast networks.  The algorithms underlying data delivery in such networks must evolve along with the networks themselves, to meet the demands of the ever-increasing number of end users.

Effectively delivering a large amount of data to a large number of users simultaneously imposes two major and seemingly contradictory demands on any system.  First, the transmission scheme must be scalable with the number of users $K$, or more precisely, the encoding operation must have low complexity in terms of the number of users.  Secondly, the system must have little to no performance degradation as the number of users increases.  That is, we wish to deliver data at rates close to channel capacity even as $K\rightarrow \infty$.

With the goal of meeting the above two demands, we study channel pre-inversion \cite{swindlehurst} and vector perturbation (VP) \cite{swindlehurst2} methods for Gaussian broadcast channels as $K\rightarrow \infty$.  Among other results, the main contribution is our \emph{max-$\SINR$ vector perturbation} (MSVP) scheme, which when combined with a low-complexity encoding algorithm has the potential to meet the demands of next-generation broadcasting networks.

\subsection{Background and Related Work}

We study a linear fading channel consisting of a transmitter with $M$ transmit antennas transmitting data to $K$ single-antenna, non-cooperating receivers, where $K\leq M$.  The basic model we consider is
\begin{equation}
{\bf y} = {\bf Hs + w}
\end{equation}
where ${\bf s}\in \mathbb{C}^M$ is an encoded data vector, ${\bf H}\in \mathbb{C}^{K\times M}$ is the channel matrix, ${\bf w}\in \mathbb{C}^K$ is additive noise, and the $i^{th}$ coordinate of ${\bf y}\in\mathbb{C}^K$ is observed by receiver $i=1,\ldots,K$.  We assume channel state information (CSI) is available at the transmitter, in which case the transmitter can write the encoded data vector as ${\bf s} = {\bf Au}$ where ${\bf u}$ is the uncoded data vector and ${\bf A}$ is a precoding matrix depending on the channel.

As was observed in \cite{swindlehurst}, the zero-forcing inverse of ${\bf H}$ is a poor choice for ${\bf A}$ when $M = K$, as the sum capacity does not scale linearly with the number of users $K$.  Instead, setting ${\bf A}$ to be a regularized inverse results in superior performance, scaling the sum capacity linearly with the number of users.  However, regularized inversion still suffers from a large gap to channel capacity when the ratio $K/M$ is close to unity.

The methods of \cite{swindlehurst} were improved upon by the \emph{vector perturbation} (VP) method of \cite{swindlehurst2}, in which a perturbation vector is added to the uncoded data vector.  Vector perturbation closes the gap to channel capacity substantially, but the transmitter is now burdened with solving a closest vector problem in an arbitrary lattice.  While algorithms such as the sphere decoder \cite{viterbo_se,damen_gamal_caire_sd} exist to tackle such problems, the complexity of finding the maximum-likelihood (ML) solution prevents VP from being scalable to a large number of users \cite{sd_complexity}.  Lattice reduction algorithms such as the LLL algorithm \cite{LLL} have been used in VP systems \cite{vp_lll,vp_lll_div}, but for very large dimensions the LLL algorithm itself can be prohibitively complex.  

Traditionally, the perturbation vector is chosen to minimize the power renormalization constant $\gamma$ (see equation (\ref{defn_gamma}) below) required at the transmitter \cite{swindlehurst2,heath_se,vp_replica,vp_revisited}.  A notable exception is \cite{wiener_se}, wherein the perturbation vector is chosen to minimize the mean square error (MSE) of the system and is shown to have superior performance compared to the `minimize $\gamma$' approach.  However, such `minimize MSE' schemes seem largely unstudied, with most authors preferring to set the precoding matrix to be the zero-forcing inverse of the channel matrix, despite poor performance for square systems at lower signal-to-noise-ratio (SNR).  The MSE of the system was also studied in \cite{vp_bd} when VP is used in conjunction with the block diagonalization technique \cite{swindlehurst3}.  VP techniques have also been studied in channels where users have multiple antennas, i.e.\ MU-MIMO channels, in \cite{vp_mumimo,vp_mumimo2,vp_mumimo3}, though we focus on the single-antenna receiver case.


\subsection{Summary of Main Contributions}

In Section \ref{sec:model} we review the regularized VP system model. In Section \ref{sec:sinr} we study the signal-to-interference-plus-noise ratio ($\SINR$) of the system and derive a useful approximation of this quantity.  In Section \ref{sec:large_mmse} we use Random Matrix Theory to predict the $\SINR$ and ergodic capacity of regularized inversion for large systems with no perturbation.  The approximation is shown to be accurate through simulations, and generalizes a theorem by the current authors for square systems ($K=M$) given in \cite{karpuk_moss}.

In Section \ref{sec:vb} we study VP and construct a scheme, which we deem \emph{max-$\SINR$ vector perturbation} (MSVP), which provably maximizes our estimate of the $\SINR$ when any regularized channel inverse is employed.  This scheme is shown to outperform the Wiener Filter VP introduced in \cite{wiener_se} which itself implicitly maximizes a different notion of $\SINR$.  We use Random Matrix Theory to estimate the performance of MSVP to within $0.5$-$1$ dB.

In Section \ref{sec:large_vb} we focus on VP for large systems, where we use the sub-ML Sorted QR method of \cite{wubben} to solve for the perturbation vector.  We show that for small $K$, the resulting performance is very close to the performance of the ML solution, and is essentially the same as that of the lattice-reduction-aided broadcast precoding of \cite{vp_lll}, even though the SQR method offers less complexity.  Lastly we show that for large $K$, MSVP outperforms the zero-forcing VP method of \cite{swindlehurst2}.  We end the paper by providing conclusions and discussing potential future work.

\subsection{Notation}

The symbols $\mathbb{Z}$, $\mathbb{R}$, and $\mathbb{C}$ denote the integers, real numbers, and complex numbers, respectively.  Capital boldface letters such as ${\bf A}$ denote matrices, and lowercase boldface letters such as ${\bf x}$ denote vectors.  We write ${\bf A}^\dagger$ for the conjugate transpose of the complex matrix ${\bf A}$, and ${\bf A}^T$ for the (non-conjugate) transpose.  If ${\bf A}$ is rectangular, its pseudo-inverse is denoted by ${\bf A}^+$.  If ${\bf A}$ is square, its trace and determinant are denoted by $\tr({\bf A})$ and $\det({\bf A})$, respectively.  The squared Frobenius norm of ${\bf A} = (a_{ij})$ is denoted by $||{\bf A}||^2_F$, and is defined by $||{\bf A}||^2_F = \tr({\bf A}^\dagger {\bf A}) = \sum_{i,j}|a_{ij}|^2$.  The identity matrix of size $K$ is denoted ${\bf I}_K$.  For any square matrix ${\bf B} = (b_{ij})$, we define a square matrix $\dg({\bf B})$ of the same size by
\begin{equation}
\dg({\bf B})_{ij} = \left\{\begin{array}{cl}
b_{ii} & \text{if $i = j$} \\
0 & \text{if $i\neq j$}
\end{array} \right.
\end{equation}
so that $\dg({\bf B})$ has the entries of ${\bf B}$ on the diagonal and zeros elsewhere.  The expectation and variance of a random variable $X$ are denoted by $\mathbb{E}(X)$ and $\text{Var}(X)$, respectively.  The Gaussian integers $\mathbb{Z}[i]$ are defined to be $\mathbb{Z}[i] = \{a+bi\ |\ a,b\in \mathbb{Z}\} \subset \mathbb{C}$ where $i^2 = -1$.

\section{System Model}\label{sec:model}

\subsection{Vector Perturbation Channel Model}

Consider the $M\times K$ MIMO channel where the transmitter has $M$ antennas and is communicating to $K\leq M$ non-cooperating users, each with a single antenna.  The intended data ${\bf u} = [u_1,\ldots,u_K]^T$ is a length $K$ column vector of information symbols (e.g.\ QAM symbols) with $u_i$ intended for receiver $i$,  normalized so that
\begin{equation}
\mathbb{E}_{\bf u}|u_i|^2 = c,\quad c= K/M.
\end{equation}
The entries of the $K\times M$ channel matrix ${\bf H}$ are i.i.d.\ circularly symmetric complex random Gaussian with variance $1/K$ per complex dimension.  The channel ${\bf H}$ is assumed to be known at the transmitter, which computes a $M\times K$ precoding matrix ${\bf A}$ and an offset perturbation vector ${\bf x}$.  The vector ${\bf x}$ is a function of both ${\bf H}$ and ${\bf u}$ and belongs to a scaled integer lattice; the precise nature of ${\bf x}$ will be made clear in the next subsection.

The transmitter computes an encoded data vector
\begin{equation}\label{defn_gamma}
{\bf s} = {\bf A({\bf u+x})}/\sqrt{\gamma},\ \text{where}\ \gamma = \mathbb{E}_{\bf u}||{\bf A({\bf u+x})}||^2/K
\end{equation}
is a power renormalization constant.  The encoded data then satisfies the power constraint $\mathbb{E}_{\bf u}||{\bf s}||^2 = K$ which allows for fair comparison when we fix $K$ and vary $M$.

The $i^{th}$ receiver observes the $i^{th}$ coordinate $y_i$ of the total length $K$ received vector 
\begin{equation}\label{basic_observation}
{\bf y} = {\bf Hs + w} = {\bf H}{\bf A({\bf u+x})}/\sqrt{\gamma} + {\bf w}
\end{equation}
from which they attempt to decode $u_i$.  Here ${\bf w} = [w_1,\ldots,w_K]^T$ is a length $K$ column vector of additive noise whose entries are i.i.d.\ circularly symmetric complex Gaussian with $\mathbb{E}_{\bf w}|w_i|^2 = \sigma^2$.  We define $\rho = 1/\sigma^2$, and will often measure system performance as a function of $\rho$ or the system size $K$.  Following convention, we set $\rho$ (dB) $= 10\log_{10}(\rho)$ and usually measure $\rho$ in dB.

\subsection{Choosing the Perturbation Vector}

The offset vector ${\bf x}$ is chosen from a scaled Gaussian integer lattice $\tau\mathbb{Z}[i]^K$ for some $\tau>0$, and may depend on both the given channel matrix ${\bf H}$ and given data vector ${\bf u}$.  Following \cite{swindlehurst2}, the scalar $\tau$ is chosen so that if the coordinates of the data vectors ${\bf u}$ are $N$-QAM constellation points, then the set
\begin{equation}
\left\{{\bf u+x}\in \mathbb{C}^K\ |\ u_i\in \text{$N$-QAM}\ \text{and}\ {\bf x}\in \tau\mathbb{Z}[i]^K\right\}
\end{equation}
is a translated lattice in $\mathbb{C}^K$.  In other words, $\tau$ is chosen so that the various translates of the set of all ${\bf u}$ are ``spaced out evenly'' throughout the Euclidean space $\mathbb{C}^K$.  One can compute easily that for unscaled, standard $N$-QAM signaling the value of $\tau$ is $2\sqrt{N}$.  For our scaling, we have
\begin{equation}\label{our_tau}
\tau = 2\sqrt{N}\frac{\sqrt{c}}{\sqrt{\frac{2}{3}(N-1)}} = \sqrt{6c\frac{N}{N-1}}
\end{equation}
where $\frac{2}{3}(N-1)$ is the average per-symbol energy of an unscaled $N$-QAM constellation.  

Following \cite{swindlehurst2}, we assume that the $i^{th}$ receiver has knowledge of $\frac{\dg({\bf HA})_{ii}}{\sqrt{\gamma}}\tau$.  The receivers model their observation as
\begin{equation}
{\bf y} = \dg({\bf HA})\frac{{\bf u+x}}{\sqrt{\gamma}} +  ({\bf HA}-\dg({\bf HA}))\frac{{\bf u+x}}{\sqrt{\gamma}}  + {\bf w}
\end{equation}
and since the $i^{th}$ receiver knows $\frac{\dg({\bf HA})_{ii}}{\sqrt{\gamma}}\tau$, they can reduce $y_i$ modulo the lattice $\frac{\dg({\bf HA})_{ii}}{\sqrt{\gamma}}\tau\mathbb{Z}[i]$ to remove the $i^{th}$ coordinate of the offset vector ${\bf x}$.    We assume that the modulo operation always decodes the offset vector ${\bf x}$ correctly, when in fact it may not if, for example, the noise vector ${\bf w}$ is very large.  However, this assumption allows for clean analysis, is pervasive in the literature, and furthermore seems to affect all VP strategies in question approximately equally.  So while our capacity plots will slightly overestimate absolute performance, they remain useful when comparing VP strategies to each other.  We restrict our VP simulations to $\rho\geq 10$ dB to mitigate the effect of this potential decoding error.


\section{Signal-to-Interference-Plus-Noise Ratio of Vector Perturbation}\label{sec:sinr}

In this section we discuss the signal-to-interference-plus-noise ratio ($\SINR$) of VP systems.  After providing the basic definition of the $\SINR$ for regularized VP systems, we show how it differs from previously considered notions of $\SINR$ for such systems (as in \cite{swindlehurst,swindlehurst2,wiener_se}), briefly explain connections with mean square error (MSE) and capacity, and provide a simple approximations of the $\SINR$ and capacity when employing a certain class of precoding matrices. 

\subsection{Basic Definition}

After successful reduction modulo the various lattices $\frac{\dg({\bf HA})_{ii}}{\sqrt{\gamma}}\tau\mathbb{Z}[i]$, the receivers model the resulting vector ${\bf y}'$ obtained from ${\bf y}$ by
\begin{equation}\label{basic_model}
{\bf y}' = \underbrace{\dg({\bf HA})\frac{{\bf u}}{\sqrt{\gamma}}}_{\text{signal}} + \underbrace{({\bf HA}-\dg({\bf HA}))\frac{{\bf u+x}}{\sqrt{\gamma}}}_{\text{interference}} + \underbrace{{\bf w}}_{\text{noise}}
\end{equation}
and treat the interference as noise when decoding.  Modeling the received signal as $\dg({\bf HA}){\bf u}/\sqrt{\gamma}$ accounts for the fact that when ${\bf A}$ is chosen to be different from the zero-forcing inverse of ${\bf H}$, the diagonal gains of the effective channel matrix ${\bf HA}$ need not be unity.  


From (\ref{basic_model}) we derive, for a fixed channel ${\bf H}$ and precoding matrix ${\bf A}$, the signal-to-interference-plus-noise ($\SINR$) ratio of the system to be
\begin{align}\label{sinr_defn}
\SINR &= \frac{\mathbb{E}_{\bf u}||\dg({\bf HA}){\bf u}||^2/\gamma}{\mathbb{E}_{\bf u}||({\bf HA}-\dg({\bf HA}))({\bf u+x})||^2/\gamma+ \mathbb{E}_{\bf w}||{\bf w}||^2} \\
&= \frac{||\dg({\bf HA})||^2_Fc}{\mathbb{E}_{\bf u}\left(||({\bf HA}-\dg({\bf HA}))({\bf u+x})||^2+ ||{\bf A({\bf u+x})}||^2\sigma^2\right)} 
\end{align}
Here we have implicitly assumed a slow fading model, wherein the channel ${\bf H}$ stays constant for a large number of transmitted data vectors ${\bf u}$.  Note that the perturbation vector ${\bf x}$ depends on ${\bf u}$, and thus there may be some correlation between the interference and signal terms in (\ref{basic_model}).  However, the interference terms is dwarfed by the noise term in practice, thus one can usually safely ignore this apparent correlation.  Secondly, the correlation between these terms at any given receiver vanishes as $K\rightarrow \infty$, as the effect of any single $u_i$ on how we choose the total perturbation vector ${\bf x}$ becomes insignificant.

We briefly point out that in \cite[Equation (25)]{swindlehurst2}, the channel model after successful reduction modulo the appropriate lattice is given by 
\begin{equation}
\widetilde{{\bf y}} = {\bf u}/\sqrt{\gamma} + ({\bf HA}-{\bf I}_K)({\bf u+x})/\sqrt{\gamma} + {\bf w}
\end{equation}
which would result in
\begin{equation}\label{wrong_sir}
\begin{aligned}
&\widetilde{\SINR} = \\
&\frac{Kc}{\mathbb{E}_{\bf u}\left(||({\bf HA} - {\bf I}_K)({\bf u+x})||^2 + ||{\bf A}({\bf u+x})||^2\sigma^2\right)}
\end{aligned}
\end{equation}
as a definition of the $\SINR$ for regularized perturbation.  However, this model overestimates the overall signal strength and therefore the capacity of the scheme, especially at low values of $\rho$ where the entries of $\dg({\bf HA})$ may be substantially smaller than unity.  We will return to this point in Section \ref{sec:vb} when we define our max-$\SINR$ vector perturbation strategy and compare it with the Wiener Filter vector perturbation method of \cite{wiener_se}, which selects the perturbation vector to maximize the $\MSE$ associated with (\ref{wrong_sir}).



\subsection{Connection to Mean Square Error and Capacity}

The connection between the $\SINR$ in equation (\ref{sinr_defn}) and the mean square error ($\MSE$) of the system is as follows.  Let us fix a data vector ${\bf u}$ and corresponding offset ${\bf x}$.  The relevant estimate of $\dg({\bf HA}){\bf u}$ at the receivers is $\hat{{\bf u}} = \sqrt{\gamma} {\bf y}'$ where ${\bf y}'$ is as in (\ref{basic_model}).  The resulting $\MSE$ for the fixed data vector ${\bf u}$ is
\begin{align}\label{mse}
\MSE_{\bf u} &= \mathbb{E}_{\bf w}||\hat{{\bf u}}-\dg({\bf HA}){\bf u}||^2   \\
&= ||({\bf HA}-\dg({\bf HA}))({\bf u+x})||^2 + ||{\bf A({\bf u+x})}||^2\sigma^2 
\end{align}
so that
\begin{equation}\label{mse2}
\SINR = \frac{||\dg({\bf HA})||^2_Fc}{\MSE},\quad \MSE = \mathbb{E}_{\bf u}(\MSE_{\bf u})
\end{equation}

The expression (\ref{basic_model}) allows one to write the resulting channel capacity for user $i$, and the average per-user capacity, for a fixed channel ${\bf H}$ as
\begin{equation}\label{cap_def}
\begin{aligned}
C_{i,{\bf H}} &= \log_2\left(1+\frac{\mathbb{E}_{\bf u}|\dg({\bf HA})_{ii}u_i|^2}{\mathbb{E}_{\bf u}|\left({\bf HA}-\dg(\bf{HA}))({\bf u+x}\right)_i|^2 + \gamma\sigma^2}\right)\\
\quad C_{\bf H} &= \frac{1}{K}\sum_{i = 1}^KC_{i,{\bf H}}
\end{aligned}
\end{equation}
respectively.  As in \cite[Equation (32)]{swindlehurst}, we make the mild assumption that the signal and interference powers are approximately uniformly distributed across all users. This allows us to approximate $C_{\bf H}\approx \log_2(1+\SINR)$ and hence the ultimate measure of capacity $\mathbb{E}_{\bf H}(C_{\bf H})$ by
\begin{equation}\label{cap_approx}
C:=\mathbb{E}_{\bf H}(C_{\bf H}) \approx \mathbb{E}_{\bf H}(\log_2(1+\SINR))
\end{equation}
The approximation (\ref{cap_approx}) is generally a good numerical estimate for the vector perturbation strategies under consideration.





\subsection{Tikhonov Pre-Inversion}

We will consider precoding matrices of the form
\begin{equation}\label{tik_def}
{\bf A} = {\bf H}_\alpha = {\bf H}^\dagger (\alpha {\bf I}_K + {\bf HH}^\dagger)^{-1}
\end{equation}
for some (small) constant $\alpha\geq0$, which is the \emph{Tikhonov inverse} of the channel matrix ${\bf H}$ with \emph{regularization parameter} $\alpha$.  When $\alpha = 0$, the Tikhonov inverse reduces to the zero-forcing inverse, which we will denote
\begin{equation}
{\bf H}_{\text{ZF}} = {\bf H}^\dagger({\bf HH}^\dagger)^{-1}
\end{equation}
When we set the regularization parameter $\alpha=\sigma^2$, we will refer to the corresponding inverse of ${\bf H}$ as the \emph{MMSE inverse}, which we will denote by
\begin{equation}
{\bf H}_{\text{MMSE}} = {\bf H}_{\sigma^2} = {\bf H}^\dagger (\sigma^2 {\bf I}_K + {\bf HH}^\dagger)^{-1}
\end{equation}
The optimal regularization parameter $\alpha$ was found in \cite{swindlehurst} to be approximately $K\sigma^2$ for square systems.  The apparent disparity with the above matrix ${\bf H}_{\text{MMSE}}$ is a consequence of how we have normalized the channel matrix and the transmit power.  We prefer the given normalization, since the regularization parameter of interest is now independent of the size of the system.

\subsection{Estimating $\SINR$ and Capacity}

When we set the precoding matrix to be a Tikhonov inverse, the resulting $\SINR$ (and thus the capacity $C$) can be estimated by a simple, compact expression.  To begin, let
\begin{equation}\label{defend}
d = \tr({\bf HH}_\alpha)/K = \tr(\dg({\bf HH}_\alpha))/K
\end{equation}
The following theorem is the basis for our approximation of the $\SINR$.
\begin{theorem}\label{Tmatrix}
For a fixed channel matrix ${\bf H}$ and Tikhonov parameter $\alpha$, define
\begin{align}
\varepsilon_1 &= ||\dg({\bf HH}_\alpha) - d{\bf I}_K||^2_F \\
\varepsilon_2 &= \mathbb{E}_{\bf u}\left(||({\bf HH}_\alpha - \dg({\bf HH}_\alpha))({\bf u+x}) ||^2\right. \nonumber\\
&\quad\quad\quad \left.- ||({\bf HH}_\alpha - d{\bf I}_K)({\bf u+x}) ||^2\right) 
\end{align}
where $d$ is as in (\ref{defend}).  Let ${\bf T}$ be any matrix such that
\begin{equation}
{\bf T}^\dagger {\bf T} = d^2{\bf I}_K -2d{\bf HH}_\alpha + {\bf HH}_\alpha {\bf H}_{\text{MMSE}}^+{\bf H}_\alpha
\end{equation}
Then we can bound the $\SINR$ by
\begin{equation}
\frac{d^2Kc}{\mathbb{E}_{\bf u}||{\bf T}({\bf u+x})||^2+\varepsilon_2} \leq \SINR\leq \frac{d^2Kc+\varepsilon_1}{\mathbb{E}_{\bf u}||{\bf T}({\bf u+x})||^2}
\end{equation}
and furthermore, we have $\lim\limits_{K\rightarrow\infty}\frac{1}{K}\mathbb{E}_{\bf H}(\varepsilon_1) = 0$.
\end{theorem}
\begin{IEEEproof}
See the Appendix.
\end{IEEEproof}

While we are unable to prove that $\lim\limits_{K\rightarrow\infty}\frac{1}{K}\mathbb{E}_{\bf H}(\varepsilon_2) = 0$, simulation results suggest that this is the case, and that even for small $K$ the quantity $\varepsilon_2$ is very small relative to the other terms in the lower bound.

We now obtain our approximations of the $\SINR$ and the capacity $C$ by ignoring the error terms $\varepsilon_1$ and $\varepsilon_2$, and setting
\begin{equation}\label{first_SINR_and_C_approx}
\widehat{\SINR} := \frac{d^2Kc}{\mathbb{E}_{\bf u}||{\bf T}({\bf u+x})||^2},\quad \widehat{C}:= \mathbb{E}_{\bf H}\log_2(1+\widehat{\SINR})
\end{equation}
where $d$ and ${\bf T}$ are as in Theorem \ref{Tmatrix}.  The above implicitly contains the approximation
\begin{equation}\label{mse_approx}
\widehat{\MSE}_{\bf u} := ||{\bf T}({\bf u+x})||^2
\end{equation}
of the mean square error for a given data vector ${\bf u}$ (and a fixed channel ${\bf H}$).

\section{Capacity of MMSE Pre-Inversion for Large Systems}\label{sec:large_mmse}

In this section we fix the offset vector ${\bf x}$ to be ${\bf x} = {\bf 0}$; that is, we are presently only concerned with the performance of linear precoding strategies with no vector perturbation.  Furthermore, we fix the precoding matrix to be ${\bf A}= {\bf H}_{\text{MMSE}}$.  The goal of this section is to obtain explicit approximations for $\mathbb{E}_{\bf H}(\SINR)$ and the capacity $C$ for MMSE pre-inversion to measure performance of large systems.

\subsection{Predicting $\SINR$ and Capacity of Large Systems}

Our strategy is to compute $\lim\limits_{K\rightarrow\infty}\mathbb{E}_{\bf H}(d)$ explicitly, and combine the result with (\ref{first_SINR_and_C_approx}) to measure large-scale system performance of MMSE pre-inversion.  This requires the following two lemmas, the first of which is an elementary simplification of our expressions for the $\SINR$ and capacity, and the second of which is a technical lemma which essentially validates replacing $d$ with $\lim\limits_{K\rightarrow\infty}\mathbb{E}_{\bf H}(d)$ in the $\SINR$ and capacity approximations.

\begin{lemma}
Suppose that ${\bf A} ={\bf H}_{\text{MMSE}}$ and that ${\bf x} = {\bf 0}$.  Then the approximations $\widehat{\SINR}$ and $\widehat{C}$ from (\ref{first_SINR_and_C_approx}) are given by
\begin{equation}
\widehat{\SINR} = \frac{d}{1-d}\quad\text{and}\quad \widehat{C} = -\mathbb{E}_{\bf H}\log_2(1-d).
\end{equation}
where $d = \tr({\bf HH}_{\text{MMSE}})/K$.
\end{lemma}
\begin{IEEEproof}
Since ${\bf x} = {\bf 0}$ and $\mathbb{E}_{\bf u}|u_i|^2 = c$, a simple calculation gives $\mathbb{E}_{\bf u}||{\bf T}({\bf u+x})||^2 = ||{\bf T}||^2_Fc$.  Since $||{\bf B}||^2_F = \tr({\bf B}^\dagger {\bf B})$ for any matrix ${\bf B}$, we see easily from (\ref{first_SINR_and_C_approx}) that
\begin{align}
\widehat{\SINR} &= \frac{d^2K}{\tr({\bf T}^\dagger {\bf T})} \\
&= \frac{d^2K}{d^2K+(1-2d)\tr({\bf HH}_{\text{MMSE}})} \\
&= \frac{d}{1-d}
\end{align}
where ${\bf T}$ is as in Theorem \ref{Tmatrix}.  The statement about $\widehat{C}$ is immediate from (\ref{first_SINR_and_C_approx}).
\end{IEEEproof}

\begin{lemma}
Let $\{X_K\}_{K = 1}^{\infty}$ be a sequence of real-valued random variables such that the following three conditions hold: (i) $0<X_K<1$ for all $K$, (ii) $0<\lim\limits_{K\rightarrow\infty}\mathbb{E}(X_K)<1$, and (iii) $\lim\limits_{K\rightarrow\infty}\text{Var}(X_K)=0$.  Then
\begin{equation}\label{exp_quotient}
\lim_{K\rightarrow \infty}\mathbb{E}\left(\frac{X_K}{1-X_K}\right) = \frac{\lim\limits_{K\rightarrow\infty}\mathbb{E}(X_K)}{1-\lim\limits_{K\rightarrow\infty}\mathbb{E}(X_K)}.
\end{equation}
\end{lemma}
\begin{IEEEproof}
See the Appendix.
\end{IEEEproof}

We can now state the main theorem of this section.

\begin{theorem}\label{main_theorem}
Let ${\bf H}$ be $K\times M$ matrix whose entries are i.i.d.\ circularly symmetric complex random Gaussian variables with variance $1/K$ per complex dimension, let ${\bf H}_\alpha = {\bf H}^\dagger (\alpha {\bf I}_K + {\bf HH}^\dagger)^{-1}$ be the Tikhonov inverse with parameter $\alpha>0$, and let $d = \tr({\bf HH}_\alpha)/K$ as in (\ref{defend}).  Then we have
\begin{equation}
\lim_{K\rightarrow\infty} \mathbb{E}_{\bf H}(d) = d(c,\alpha)
\end{equation}
where
\begin{equation}
d(c,\alpha):=\frac{1+c+c\alpha -\sqrt{1+2c(-1+\alpha) + c^2(1+\alpha)^2}}{2c}.
\end{equation}
Furthermore, if we fix ${\bf A} = {\bf H}_{\text{MMSE}}$, then
\begin{align}
\lim_{K\rightarrow\infty}\mathbb{E}_{\bf H}(\widehat{\SINR}) &= \frac{d(c,\sigma^2)}{1-d(c,\sigma^2)}
\end{align}
and
\begin{align}
\lim_{K\rightarrow\infty} \widehat{C}&\geq -\log_2(1-d(c,\sigma^2))
\end{align}
for systems which employ MMSE pre-inversion with no vector perturbation.
\end{theorem}
\begin{IEEEproof}  See the Appendix.
\end{IEEEproof}

Theorem \ref{main_theorem} provides the approximations
\begin{align}\label{plain_tik_approx}
\mathbb{E}_{\bf H}(\SINR)&\approx E_{\text{MMSE}}(c,\sigma^2) := \frac{d(c,\sigma^2)}{1-d(c,\sigma^2)},\\
C&\approx -\log_2(1-d(c,\sigma^2))
\end{align}
for large systems which employ MMSE pre-inversion with $\alpha = \sigma^2$ and no vector perturbation.

\subsection{Simulation Results}

In this subsection we collect simulation results which study the accuracy and predictive ability of the approximation (\ref{plain_tik_approx}), for channel pre-inversion with ${\bf A} = {\bf H}_{\text{MMSE}}$ and no vector perturbation.

\subsubsection{Approximating $\SINR$}

\begin{figure}[h!]
\centering
\includegraphics[width=.40\textwidth]{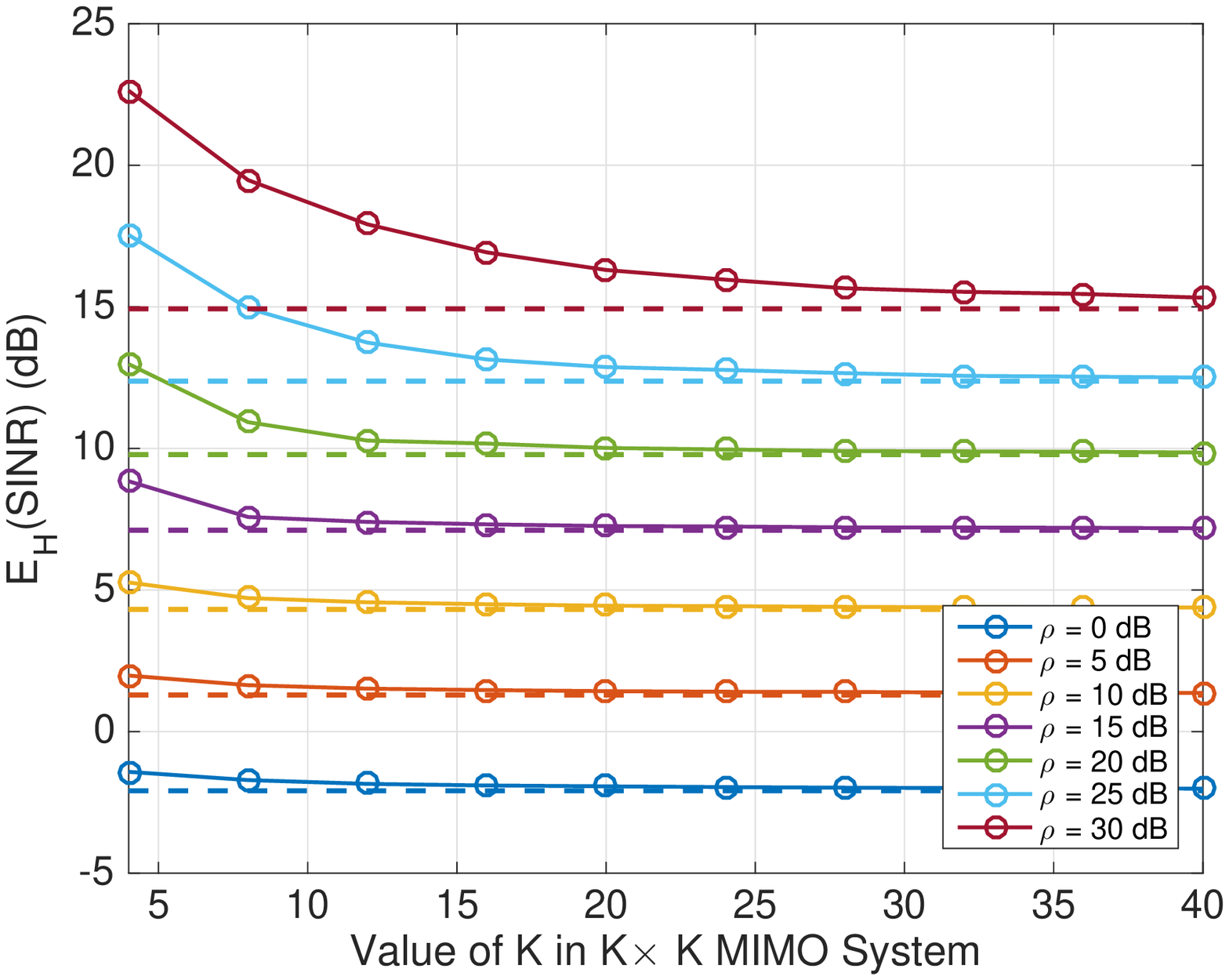} \hfill
\includegraphics[width=.40\textwidth]{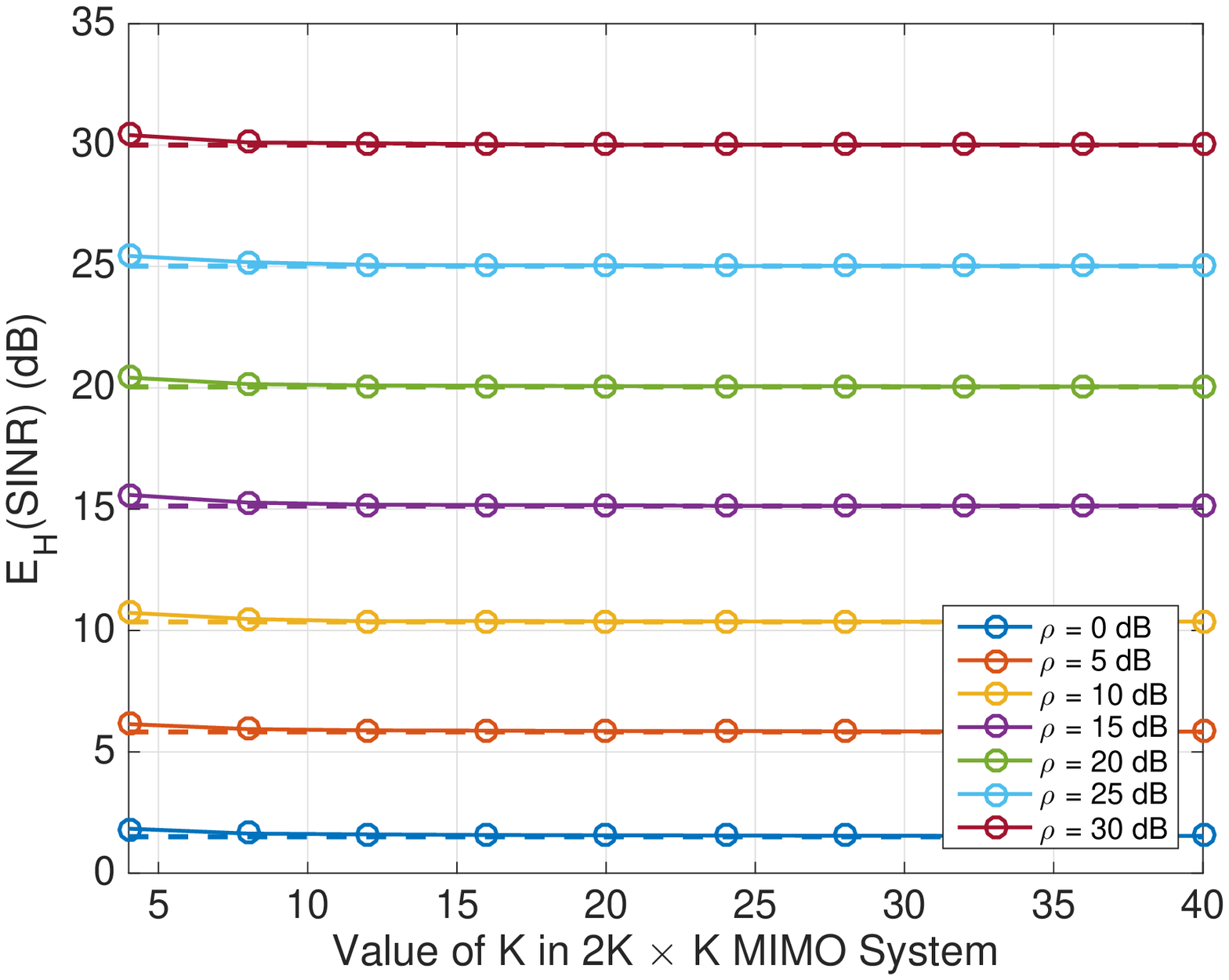}
\caption{On top, $\mathbb{E}_{\bf H}(\SINR)$ for $K\times K$ systems $(c = 1)$ employing MMSE pre-inversion, for various values of $\rho = 1/\sigma^2$.  On bottom, the same plot for $2K\times K$ systems.}\label{SINR_KbyK_tik}
\end{figure}

In Fig.\ \ref{SINR_KbyK_tik} we plot $\mathbb{E}_{\bf H}(\SINR)$ as a function of $K$ for systems with $M = K$ (top), and $M = 2K$ (bottom), for various values of $\rho = 1/\sigma^2$.  In both plots, the solid curves represent experimentally measured values of $\mathbb{E}_{\bf H}(\SINR)$, and the dashed lines the corresponding values of $E_{\text{MMSE}}(c,\sigma^2)$.  We see that $E_{\text{MMSE}}(c,\sigma^2)$ predicts the limiting value of $\mathbb{E}_{\bf H}(\SINR)$ very well, for all values of $\rho$.  On the other hand, note that the error introduced by applying the large-$K$ limit to small-$K$ systems may be non-negligible when $c = 1$.  





\subsubsection{Approximating Capacity}

\begin{figure}[h!]
\centering
\includegraphics[width=.40\textwidth]{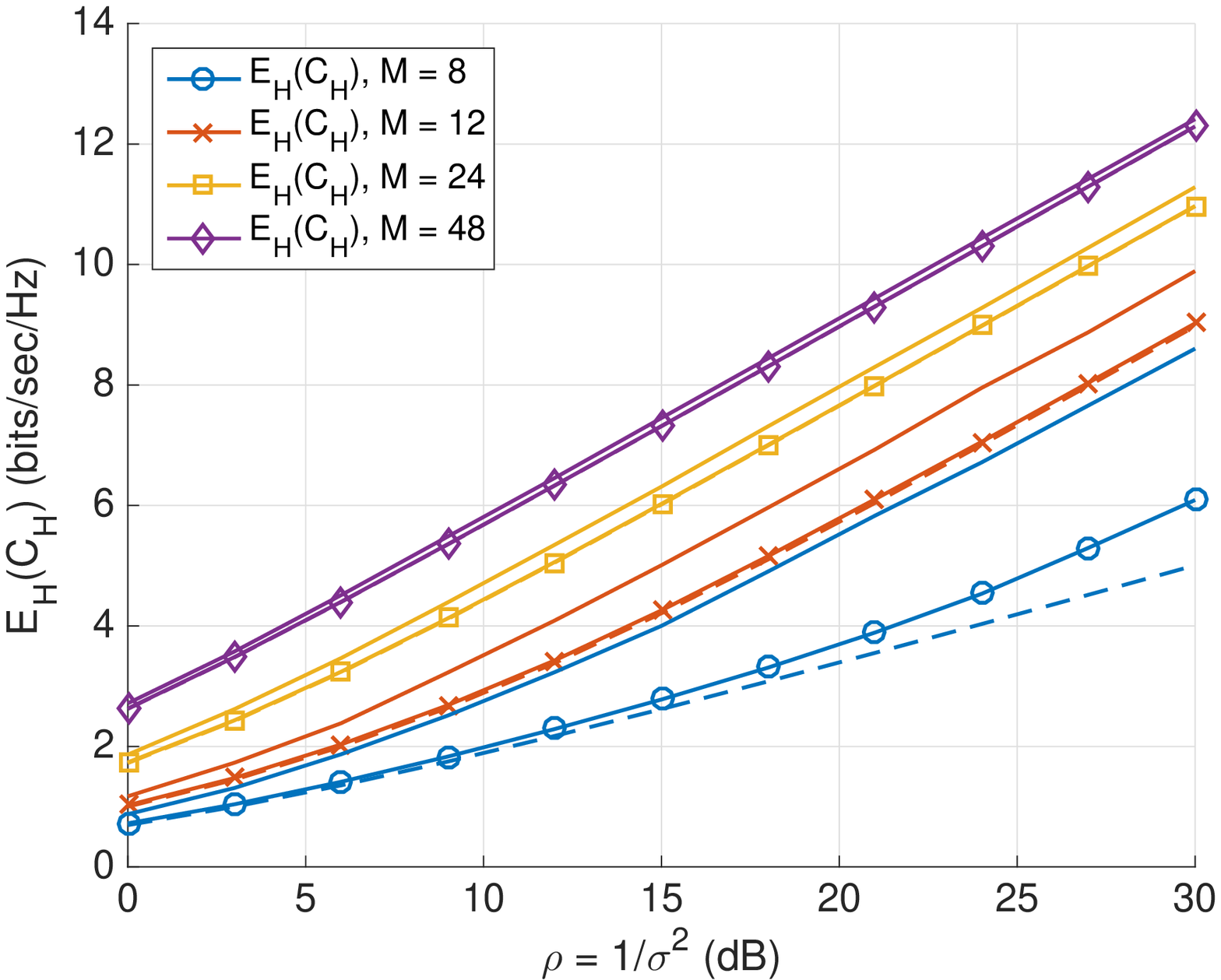}
\hfill
\includegraphics[width=.40\textwidth]{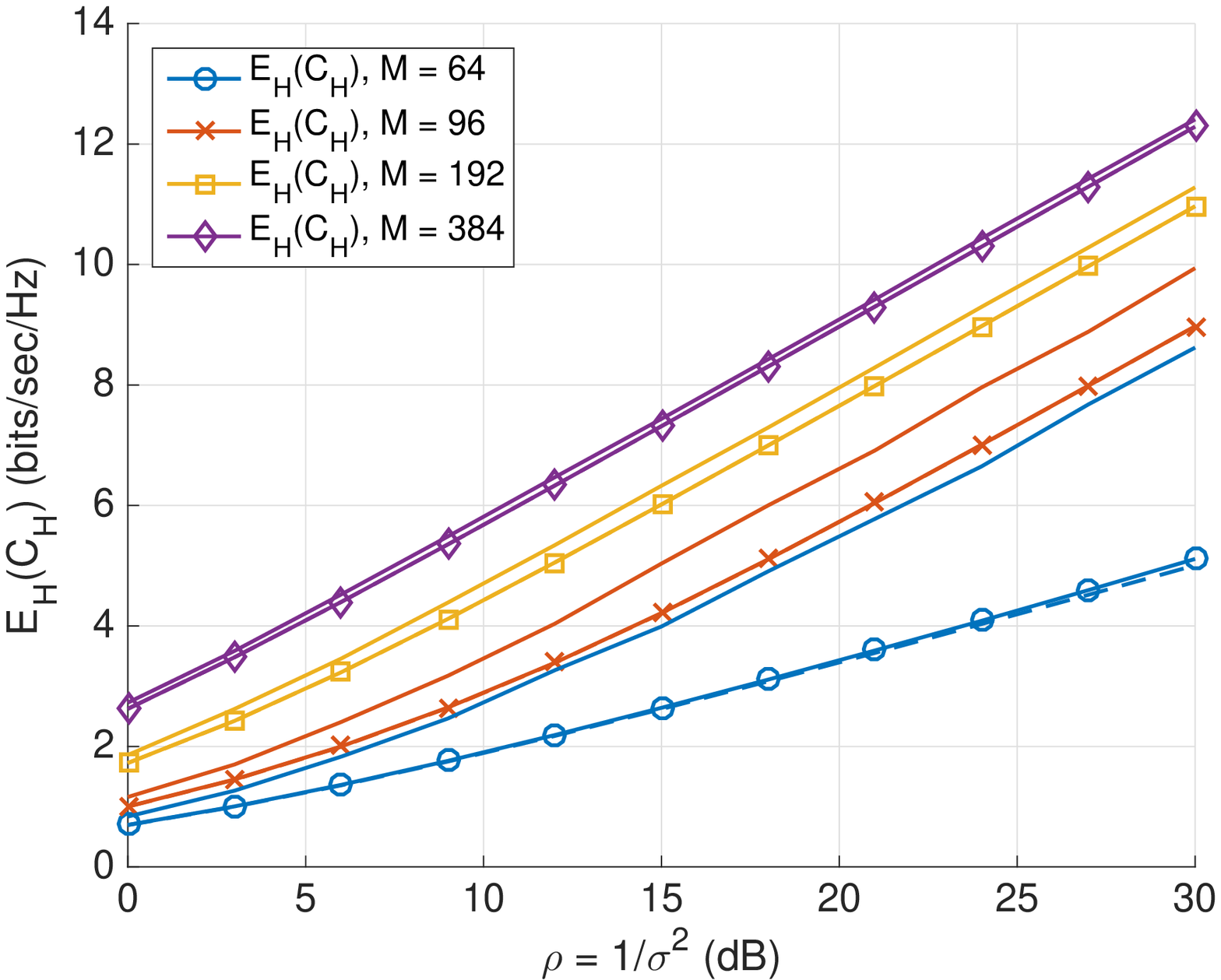}
\caption{Capacity of MMSE pre-inversion for $K = 8$ (top) and $K = 64$ (bottom).  The marked lines are the experimentally computed capacity $C$ of MMSE pre-inversion, the dashed lines represent our approximation $-\log_2(1-d(c,\sigma^2))$ of this quantity, and the solid lines the channel capacity.}\label{tik_approx}
\end{figure}

In Fig.\ \ref{tik_approx} we plot the capacity $C = \mathbb{E}_{\bf H}(C_{\bf H})$ as a function of $\rho$ for $K = 8$ (top) and $K = 64$ (bottom), for $c = 1$, $2/3$, $1/3$, $1/6$.  The solid marked lines are the experimentally measured values of $C$, the dashed lines are the values of the approximation $-\log_2(1-d(c,\sigma^2))$ of $C$ from (\ref{plain_tik_approx}), and the solid unmarked lines are the channel capacity, computed numerically using the results of \cite{tse_vis_broadcast_cap}.  We see that the approximation (\ref{plain_tik_approx}) of the capacity is very good in all scenarios, except for small square systems when $\rho$ is large.

%

\subsection{Qualitative Behavior of Capacity of MMSE Pre-Inversion}

To study the qualitative behavior of the capacity of Tikhonov pre-inversion with ${\bf A}={\bf H}_{\text{MMSE}}$, we expand $-\log_2(1-d(c,\sigma^2))$ in a Taylor series as $\sigma^2\rightarrow0$ to obtain (recall that $\rho = 1/\sigma^2$)
\begin{equation}\label{big_approx}
\begin{aligned}
C&\gtrsim-\log_2(1-d(c,\sigma^2)) \\
&= \left\{\begin{array}{cl}
\log_2(\rho) +\log_2(\frac{1-c}{c})+O(1) & c<1 \\
\frac{1}{2}\log_2(\rho) +O(1) & c=1.
\end{array}
\right.
\end{aligned}
\end{equation}
The experimental results of the previous subsection show that the leading term of the series approximates $C$ quite well for $\rho > 15$ dB or so.  If one accepts that $-\log_2(1-d(c,\sigma^2))$ is an accurate predictor of $C$ for large $K$, then the above shows that the qualitative performance of square and non-square systems is different.


\section{max-$\SINR$ Vector Perturbation}\label{sec:vb}

In this and all subsequent sections we turn our attention towards schemes which employ non-trivial vector perturbation, that is, choose ${\bf x}\in \tau\mathbb{Z}[i]^K$ according to some algorithm which is intended to optimize system performance.  In \cite{swindlehurst2} and nearly all subsequent literature, the authors fix the precoding matrix to be ${\bf A}= {\bf H}_\alpha$ for $\alpha\geq 0$, and for a fixed data vector ${\bf u}$ choose the offset vector ${\bf x}$ to be
\begin{equation}\label{bad_x}
{\bf x} = \underset{{\bf x}'\in \tau\mathbb{Z}[i]^K}{\argmin}\ \gamma \
= \underset{{\bf x}'\in \tau\mathbb{Z}[i]^K}{\argmin}\ ||{\bf A}({\bf u} + {\bf x}')||^2
\end{equation}
which is a closest-vector problem in a lattice and hence solvable with a sphere decoder.  


\subsection{Max-$\SINR$ Vector Perturbation}\label{sec:max_sinr_vp}

Rather than choosing ${\bf x}$ to minimize $\gamma$, we instead choose ${\bf x}$ to minimize the mean square error of the system.  Specifically, for a fixed channel matrix ${\bf H}$ and a fixed data vector ${\bf u}$, we choose the perturbation vector ${\bf x}$ according to
\begin{equation}\label{best_x}
{\bf x} = \underset{{\bf x}'\in\tau\mathbb{Z}[i]^K}{\argmin}\ \MSE_{\bf u} = \underset{{\bf x}'\in \tau\mathbb{Z}[i]^K}{\argmin}\ ||{\bf T}({\bf u+x}')||^2 
\end{equation}
where ${\bf T}$ satisfies ${\bf T}^\dagger {\bf T} = d^2{\bf I}_K - 2d{\bf HH}_\alpha + {\bf HH}_\alpha {\bf H}_{\text{MMSE}}^+ {\bf H}_\alpha $ and $d = \tr({\bf HH}_\alpha)/K$.  Note that this provides a VP strategy for any regularization parameter $\alpha \geq 0$ whatsoever, not just the MMSE parameter $\alpha = \sigma^2$.  We will refer to this strategy as \emph{max-$\SINR$ vector perturbation}, or MSVP.

We emphasize that this is \emph{not} the Wiener Filter VP strategy (WFVP) of \cite{wiener_se}, which chooses the offset vector ${\bf x}$ to minimize $||{\bf L}({\bf u+x'})||^2$, where ${\bf L^\dagger L} = (\sigma^2 {\bf I}_K + {\bf HH}^\dagger)^{-1}$.  An argument similar to the proof of our Theorem 1 shows that $||{\bf L}({\bf u+x})||^2$ is the denominator of the alternative expression (\ref{wrong_sir}) of the $\SINR$.  Thus WFVP attempts to maximize the $\SINR$, but does not account for the diagonal entries of ${\bf HH}_\alpha$ being less than unity.  

\begin{figure}[h!]
\centering
\includegraphics[width=.40\textwidth]{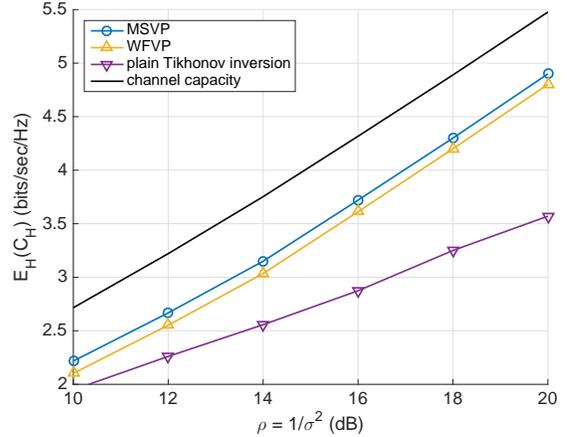}
\caption{Capacity of the WFVP strategy of \cite{wiener_se} and the MSVP defined by (\ref{best_x}), for a system with $K = M = 12$.}\label{VP_caps_K12M12_MSVP_vs_WFVP}
\end{figure}

To demonstrate the improvement offered by our MSVP method over the WFVP strategy of \cite{wiener_se}, we plot the capacity $C = \mathbb{E}_{\bf H}(C_{\bf H})$ as defined in (\ref{cap_approx}) of both schemes in Fig.\ \ref{VP_caps_K12M12_MSVP_vs_WFVP} for systems with $K = M = 12$ employing $16$-QAM modulation.  We see a consistent gain of approximately $0.5$ dB over the WFVP strategy.


\subsection{Estimating the Performance of max-$\SINR$ Vector Perturbation}

In this subsection we will estimate the performance of MSVP when using the regularization parameter $\alpha = \sigma^2$.  The regularization parameter $\alpha = \sigma^2$ is not known to be the optimal regularization parameter for the MSVP strategy, and without knowledge of the optimal $\alpha$ choosing $\alpha = \sigma^2$ is simply convenient.  The main result of this subsection is an approximation of $\mathbb{E}_{\bf H}(\SINR)$ which can be used to predict system performance to within about $0.5$-$1$ dB, which we demonstrate through numerous simulations.

The results of \cite{heath_se}, specifically \cite[Lemma 1 and Corollary 1]{heath_se}, estimate the power renormalization constant $\gamma$ for the precoding matrix ${\bf A}= {\bf H}_{\text{ZF}}$; Jensen's Inequality can then be used to estimate the expected $\SINR$ for such a `zero-forcing' strategy.  However the MSVP strategy, which sets ${\bf A} = {\bf H}_{\text{MMSE}}$ and chooses ${\bf x}$ according to (\ref{best_x}), has qualitatively different performance at lower values of $\rho$ when compared to the `zero-forcing' strategy.  Thus a new predictor of performance is required.

While the estimate of $\gamma$ for the `zero-forcing vector perturbation' method of \cite{heath_se} does not provide a useful predictor for the $\SINR$ of max-$\SINR$ vector perturbation, the general strategy therein remains applicable.  In particular, we let
\begin{equation}
\begin{aligned}
\mathcal{H}_K &= \text{ hypercube in $\mathbb{C}^K$, with side length} \\
&\lim_{N\rightarrow \infty}\tau = \lim_{N\rightarrow \infty}\sqrt{6c\frac{N}{N-1}} = \sqrt{6c}
\end{aligned}
\end{equation}
and we consider data vectors ${\bf u}$ chosen from the uniform distribution on $\mathcal{H}_K$.  Heuristically, we are approximating the discrete $N$-QAM distribution by the uniform input distribution on the minimal hypercube surrounding the constellation as $N\rightarrow \infty$.  One can check that our energy constraint is preserved, in other words that for such uniform inputs ${\bf u}$, we have
\begin{equation}
\mathbb{E}_{\bf u}||{\bf u}||^2 = \frac{1}{\vol(\mathcal{H}_K)}\int_{\mathcal{H}_K}||{\bf t}||^2 d{\bf t} = Kc
\end{equation}
and hence $\mathbb{E}_{\bf u}|u_i|^2 = c$ since the entries of the data vector ${\bf u}$ are assumed i.i.d.  

Recall from (\ref{mse2}) and (\ref{mse_approx}) that for a fixed channel matrix ${\bf H}$, our estimate $\widehat{\MSE}$ of the mean square error of the system is given by
\begin{equation}
\widehat{\MSE} = \mathbb{E}_{\bf u}||{\bf T}({\bf u+x})||^2
\end{equation}
where ${\bf T}$ is as in Theorem \ref{Tmatrix}.  The below proposition, modeled on \cite[Lemma 1]{heath_se}, will allow us to estimate $\mathbb{E}_{\bf H}(\widehat{\MSE})$ and thus provide a useful estimate of $\mathbb{E}_{\bf H}(\SINR)$.

\begin{prop}\label{est_max_sinr1}
Suppose an $M\times K$ MIMO system employs MSVP with a fixed channel matrix ${\bf H}$ and data vectors ${\bf u}$ uniform on $\mathcal{H}_K$, with any regularization parameter $\alpha\geq 0$.  Then we have
\begin{equation}
\widehat{\MSE} \geq \frac{6c}{\pi}\frac{K(K!)^{1/K}}{K+1}\det({\bf T}^\dagger {\bf T})^{1/K}
\end{equation}
where ${\bf T}$ is as in Theorem \ref{Tmatrix}.
\end{prop}
\begin{IEEEproof}  The proof is the same as for \cite[Lemma 1]{heath_se}, wherein $\widehat{\MSE}$ is expressed as the second moment of the Voronoi cell of the lattice generated by ${\bf T}$ in $\mathbb{C}^K$, and is related to the second moment of the unit sphere in $\mathbb{R}^{2K}$.  We omit further details.
\end{IEEEproof}

To apply the above result to approximate the expected $\SINR$ of the system, we will need to estimate $\mathbb{E}_{\bf H}(\det({\bf T}^\dagger {\bf T}))$.  To that end, we recall a result from Random Matrix Theory.  Let ${\bf W} = K{\bf HH}^\dagger$ be taken from the complex Wishart distribution $\mathcal{W}_K(M,{\bf I}_K)$ \cite[Section 2.1.3]{verdu_rmt}, and let $\beta$ be constant with respect to ${\bf H}$.  Then \cite[Theorem 2.13]{verdu_rmt} states that
\begin{equation}\label{big_rmt}
\mathbb{E}_{\bf H}(\det({\bf I}_K+\beta {\bf W})) = \sum_{i = 0}^K\binom{K}{i}\frac{M!}{(M-i)!}\beta^i.
\end{equation}
When $\alpha = \sigma^2$ we can rewrite ${\bf T}^\dagger {\bf T}$ as a product of matrices of the form ${\bf I}_K + \beta {\bf W}$ and then apply the above result to estimate $\mathbb{E}_{\bf H}(\det({\bf T}^\dagger {\bf T}))$.  Straightforward computation gives
\begin{equation}
{\bf T}^\dagger {\bf T} 
= d^2\left({\bf I}_K + \left(\frac{1-d}{d}\right)^2\frac{1}{K\sigma^2}{\bf W}\right)\left({\bf I}_K+\frac{1}{K\sigma^2}{\bf W}\right)^{-1} \nonumber
\end{equation}
Replacing $d$ with its large-$K$ limit $d(c,\sigma^2)$, again approximating the expectation of a ratio by the ratio of the expectations, and using (\ref{big_rmt}) we arrive at
\begin{equation}
\begin{aligned}
\mathbb{E}_{\bf H}(\det({\bf T}^\dagger {\bf T})) &\approx d(c,\sigma^2)^{2K}\frac{\mathbb{E}_{\bf H}(\det({\bf I}_K+\beta_1W))}{\mathbb{E}_{\bf H}(\det({\bf I}_K+\beta_2W))} \\
&= d(c,\sigma^2)^{2K}\frac{\sum_{i = 0}^K\binom{K}{i}\frac{M!}{(M-i)!}\beta_1^i}{\sum_{i = 0}^K\binom{K}{i}\frac{M!}{(M-i)!}\beta_2^i}
\end{aligned}
\end{equation}
where
\begin{equation}\label{gammas}
\beta_1 = \left(\frac{1-d(c,\sigma^2)}{d(c,\sigma^2)}\right)^2\frac{1}{K\sigma^2},\quad \beta_2 = \frac{1}{K\sigma^2}.
\end{equation}

We can complete our approximation of $\mathbb{E}_{\bf H}(\SINR)$ by performing the following series of approximations:
\begin{equation}\label{loosey_goosey}
\begin{aligned}
\mathbb{E}_{\bf H}(\SINR) 
&\approx \frac{d(c,\sigma^2)^2Kc}{\mathbb{E}_{\bf H}(\widehat{\MSE})} \\
&\leq \frac{d(c,\sigma^2)^2Kc}{\frac{6c}{\pi}\frac{K(K!)^{1/K}}{K+1}\mathbb{E}_{\bf H}(\det({\bf T}^\dagger {\bf T})^{1/K})} \\
&\approx \frac{\pi}{6}\frac{K+1}{(K!)^{1/K}}\frac{d(c,\sigma^2)^2}{(\mathbb{E}_{\bf H}\det({\bf T}^\dagger {\bf T}))^{1/K}} \\
&\approx \frac{\pi}{6}\frac{K+1}{(K!)^{1/K}} \left[\frac{\sum_{i = 0}^K\binom{K}{i}\frac{M!}{(M-i)!}\beta_2^i}{\sum_{i = 0}^K\binom{K}{i}\frac{M!}{(M-i)!}\beta_1^i}\right]^{1/K}
\end{aligned}
\end{equation}
where $\beta_1$ and $\beta_2$ are as in equation (\ref{gammas}).  We now have
\begin{equation}\label{max_sinr_approx}
\mathbb{E}_{\bf H}(\SINR) \approx E_{\text{vp}}(K,M,\sigma^2)
\end{equation}
where
\begin{equation}
E_{\text{vp}}(K,M,\sigma^2):= \frac{\pi}{6}\frac{K+1}{(K!)^{1/K}} \left[\frac{\sum_{i = 0}^K\binom{K}{i}\frac{M!}{(M-i)!}\beta_2^i}{\sum_{i = 0}^K\binom{K}{i}\frac{M!}{(M-i)!}\beta_1^i}\right]^{1/K} 
\end{equation}
for max-$\SINR$ vector perturbation with  $\alpha = \sigma^2$.  The approximations in (\ref{loosey_goosey}) all stem from replacing a quantity by its large-$K$ limit, or from an application of Jensen's Inequality.  We omit further details in favor of demonstrating the validity of the approximation through simulations.

\subsection{Simulation Results}

In this subsection we empirically demonstrate the accuracy of (\ref{max_sinr_approx}) for MSVP with $\alpha = \sigma^2$.  We fix the signaling alphabet to be a $16$-QAM constellation for all experiments.  The solid curves represent experimentally measured values of $\mathbb{E}_{\bf H}(\SINR)$ or the capacity $C$, and the dashed curves the resulting approximations.

\subsubsection{$\SINR$ as a function of $K$}  
\begin{figure}[h!]
\centering
\includegraphics[width=.40\textwidth]{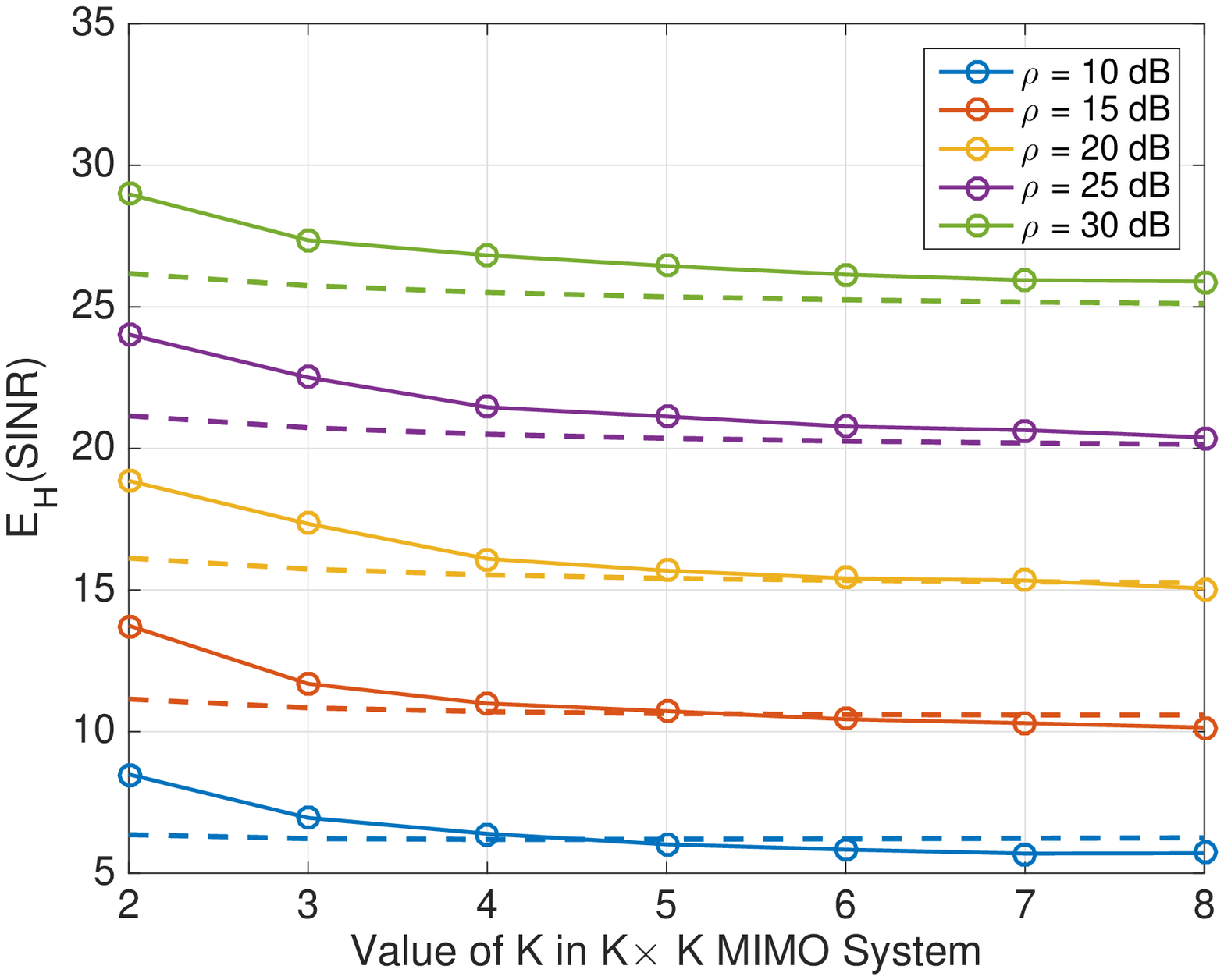} \hfill
\includegraphics[width=.40\textwidth]{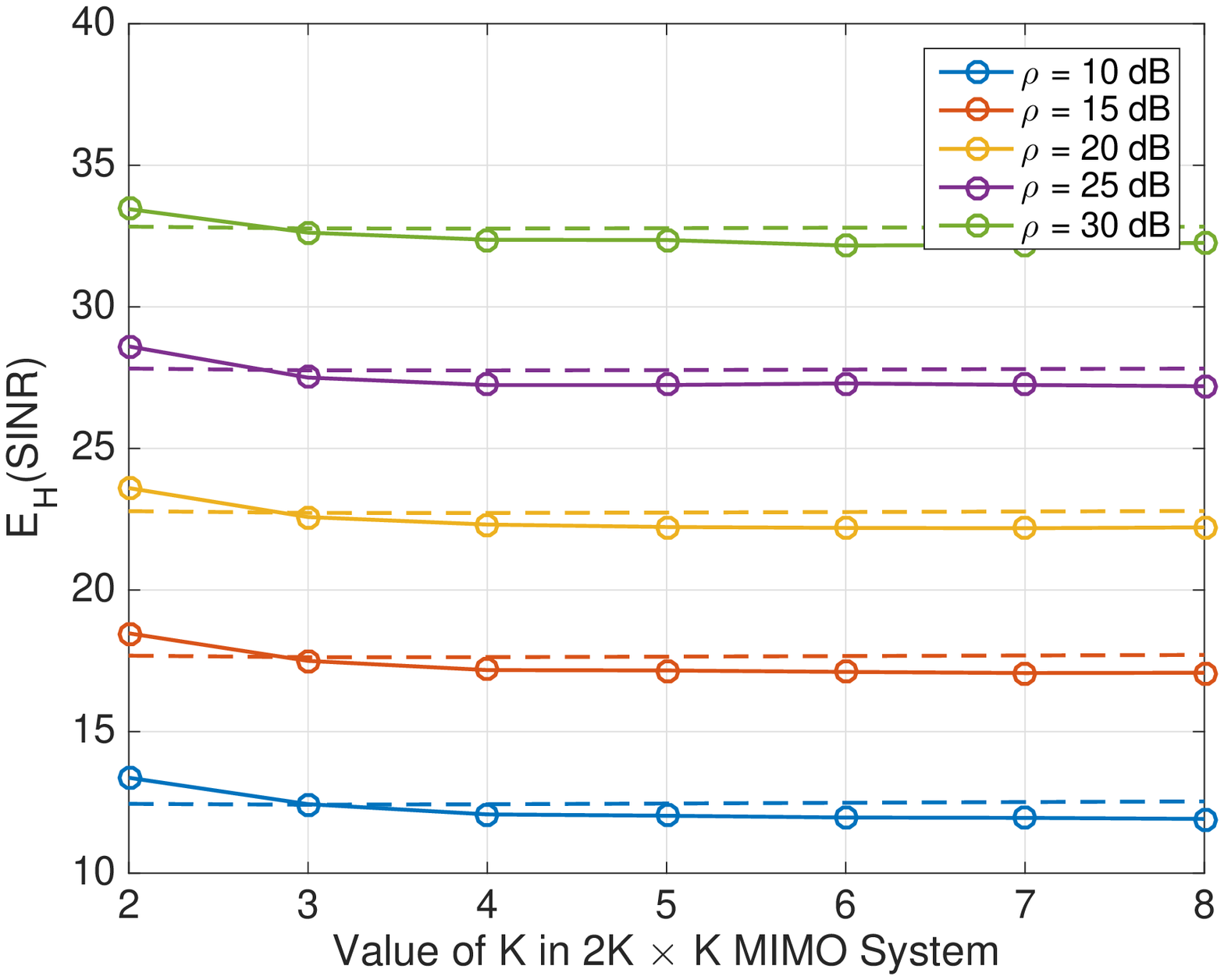}
\caption{On top, $\mathbb{E}_{\bf H}(\SINR)$ for a $K\times K$ system $(c = 1)$ employing max-$\SINR$ vector perturbation with $\alpha = \sigma^2$, for various values of $\rho = 1/\sigma^2$.  On bottom, the same plot for $2K\times K$ systems.}\label{SINR_fn_of_K_maxSINR_vp_c1}
\end{figure}

We study (\ref{max_sinr_approx}) in Fig.\ \ref{SINR_fn_of_K_maxSINR_vp_c1} for $M=K$ (top) and $M = 2K$ (bottom).  We see that the approximation is accurate to within about 1 dB when $K\geq 4$.  Implicit in our approximation (\ref{max_sinr_approx}) is (\ref{plain_tik_approx}) which essentially replaces the value $d$ by its large-$K$ limit, hence one should expect (\ref{max_sinr_approx}) to also be more accurate for larger $K$.  

\subsubsection{Approximating Capacity}  
\begin{figure}[h!]
\centering
\includegraphics[width=.40\textwidth]{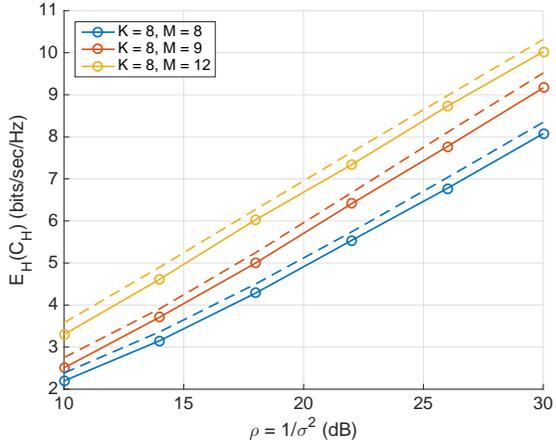} \hfill
\caption{The experimentally measured capacity $C$ of MSVP for systems with $K = 8$ and $M = 8$, $9$, and $12$, versus the approximation $\log_2(1+E_{\text{vp}}(K,M,\sigma^2))$ of this quantity.}\label{VP_caps_K8M8_MSVP_vs_approx}
\end{figure}

In Fig.\ \ref{VP_caps_K8M8_MSVP_vs_approx} we plot, for $K = 8$ and $c = 1$, $8/9$, and $4/5$, the ergodic capacity $C$ of max-$\SINR$ vector perturbation as well as the corresponding estimate obtained by combining (\ref{max_sinr_approx}) and (\ref{cap_approx}) to obtain the approximation $C \approx \log_2(1+E_{\text{vp}}(K,M,\sigma^2))$.  Again, we see that this estimate predicts the expected capacity well. 

\section{max-$\SINR$ Vector Perturbation for Large Systems}\label{sec:large_vb}

Vector perturbation offers large benefits over channel inversion, but computing the optimal offset vector ${\bf x}$ in (\ref{best_x}) may be prohibitively complex for large $K$.  In \cite{vp_lll} the authors used the LLL lattice-reduction algorithm to achieve this goal, but for very large $K$ this reduction itself can be prohibitively complex.  We will use a method with even smaller complexity, which we show approaches the performance of the ML solution for small $K$, and slightly outperforms LLL-based precoding for large $K$.  For all experiments $16$-QAM modulation was used.

\subsection{Sorted QR max-$\SINR$ Vector Perturbation}\label{sqr}

We employ the \emph{Sorted QR Precoding} (SQR) method of \cite{wubben}, a sub-ML algorithm for decoding space-time codes which can be summarized as follows.  For our purposes it suffices to consider the problem of computing
\begin{equation}\label{basic_sd}
{\bf x} = \underset{{\bf x}'}{\argmin}\ ||{\bf y} - {\bf Tx}'||^2
\end{equation}
for a square $K\times K$ matrix ${\bf T}$ where ${\bf x}'$ ranges over an integer lattice.  The SQRP algorithm is a modified Gram-Schmidt procedure which decomposes the $K\times K$ matrix ${\bf T}$ as a product
\begin{equation}\label{ldecomp}
{\bf T} = {\bf QRP}
\end{equation}
where ${\bf Q}$ is $K\times K$ unitary, ${\bf R} = (r_{ij})_{1\leq i,j\leq K}$ is $K\times K$ upper-right triangular, and ${\bf P}$ is a $K\times K$ permutation matrix, to attempt to maximize the diagonal entries $r_{ii}$ of ${\bf R}$, in order as $i = K,\ldots,1$.  Substituting into (\ref{basic_sd}) we obtain
\begin{align}
\underset{{\bf x}'}{\argmin}||{\bf y}-{\bf Tx}'||^2 &= \underset{{\bf x}'}{\argmin}||{\bf y} - {\bf QRPx}'||^2 \\
&= \underset{{\bf z}}{\argmin}||\tilde{{\bf y}} - {\bf Rz}||^2 \label{ldecomp_z}
\end{align}
where $\tilde{{\bf y}} = {\bf Q}^\dagger {\bf y}$ and ${\bf z} = {\bf Px}'$.

Let us recall the definition of the \emph{Babai point} ${\bf z}^B = [z_1^B,\ldots,z_K^B]^T$, an estimate of the solution to (\ref{ldecomp_z}) given recursively by
\begin{align}
c_K &= \tilde{y}_K/r_{KK},\quad z^B_K = [c_K] \label{babai1}\\
c_i &= (\tilde{y}_i - \sum_{j = i+1}^Kr_{ij}z^B_j)/r_{ii},\quad z^B_i = [c_i],\\
 &\quad\ \text{for $i = K-1,\ldots,1$}\nonumber \label{babai2}
\end{align}
where $[\cdot]$ denotes rounding to the nearest element of the underlying per-coordinate constellation.  The final estimate of the ML solution ${\bf x}$ is obtained by computing ${\bf P}^{-1}{\bf z}^B$.   The modified SQR algorithm of \cite{wubben} increases the probability that ${\bf P}^{-1}{\bf z}^B$ is the ML solution to (\ref{basic_sd}).  We refer to \cite{wubben} for further details.

To apply this algorithm to the VP procedure, we rewrite the argmin problem (\ref{best_x}) as
\begin{align}
{\bf x} = \underset{{\bf x}'\in \mathbb{Z}[i]^K}{\argmin}\ ||{\bf T}({\bf u} + {\bf x}')||^2 = -\underset{{\bf x}'\in \mathbb{Z}[i]^K}{\argmin}\ ||{\bf y} - {\bf Tx}'||^2
\end{align}
where ${\bf y} = {\bf Tu}$ is the `received' vector.  We then apply the decomposition (\ref{ldecomp}) and compute the estimate of ${\bf x}$, namely ${\bf P}^{-1}{\bf z}^B$, as above.  We refer to the process as \emph{Sorted QR vector perturbation}, or just \emph{SQR vector perturbation}.

\subsection{Comparison with ML and Lattice-Reduction-Aided Broadcast Precoding}

\begin{figure}[h!]
\centering
\includegraphics[width=.40\textwidth]{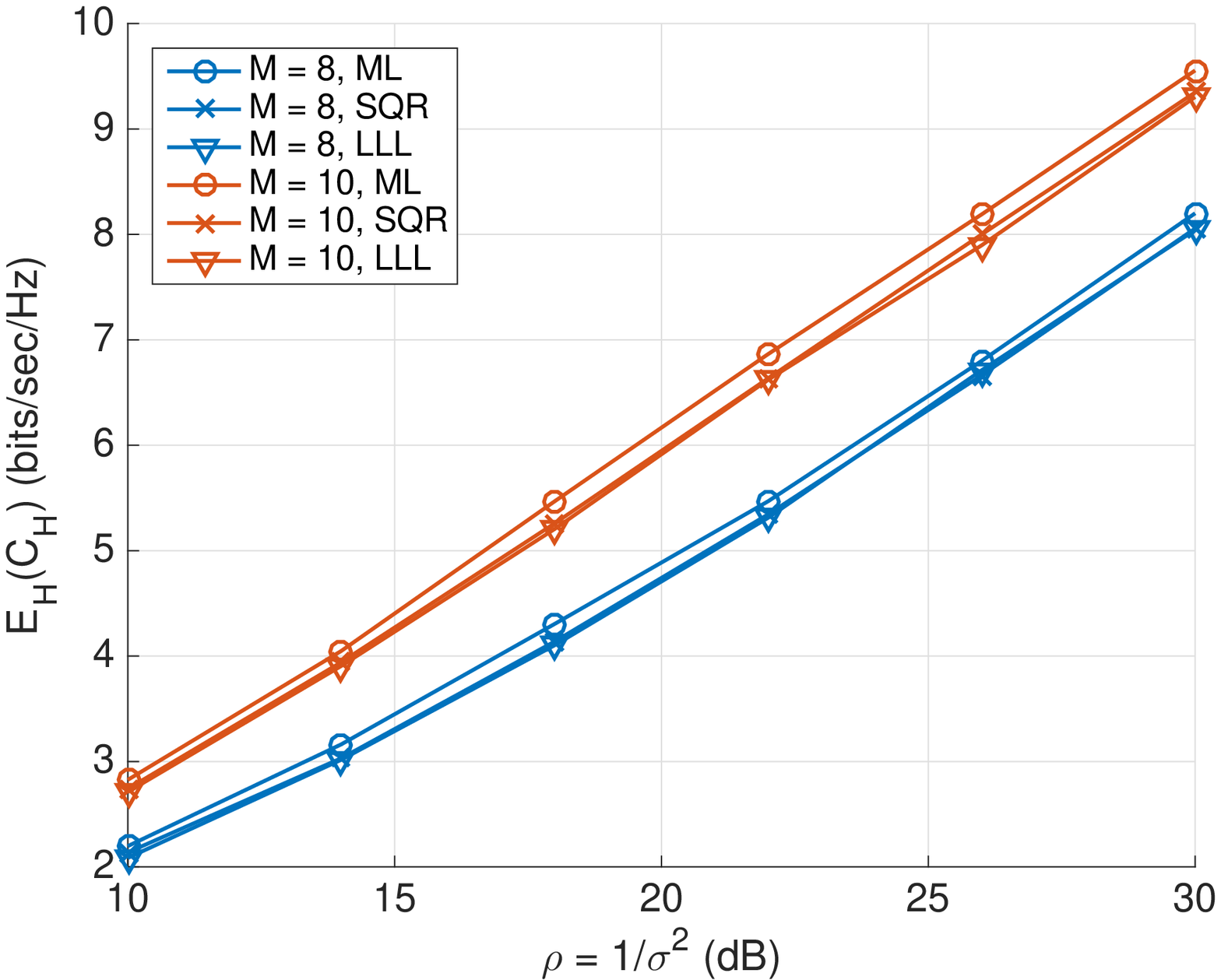} \hfill
\includegraphics[width=.40\textwidth]{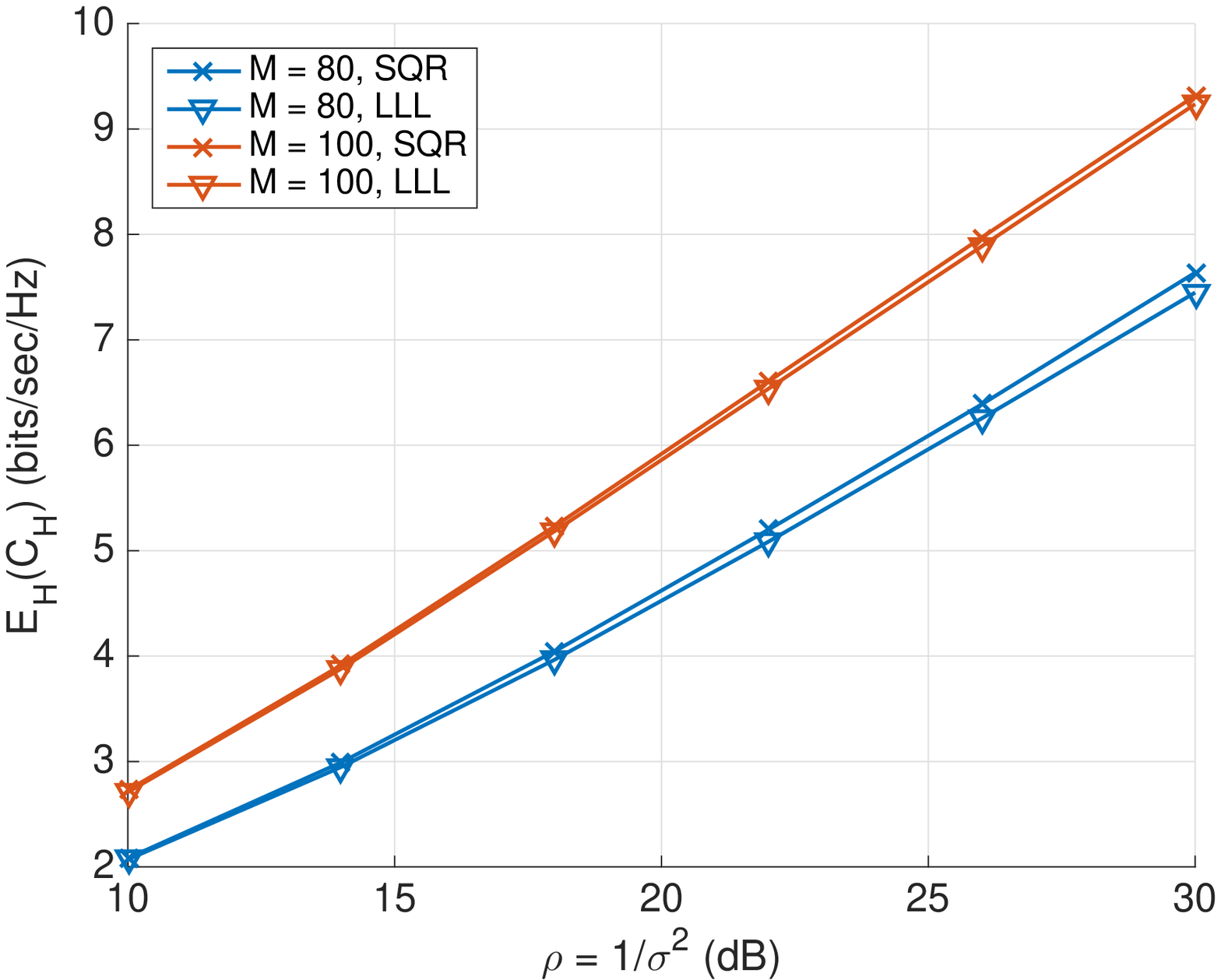}
\caption{On top, capacity of MSVP systems with $K = 8$ and $c = 1$, $4/5$ employing ML,  SQR, and LLL methods for computing the perturbation vector ${\bf x}$.  On bottom, the same plot for $K = 80$, omitting the ML strategy.}\label{ML_vs_SQR_vs_LLL}
\end{figure}

In the top plot of Fig.\ \ref{ML_vs_SQR_vs_LLL} we compare SQR MSVP for $K = 8$ and $c = 1$, $4/5$ to ML MSVP wherein (\ref{best_x}) is solved using a sphere decoder.  We also plot the performance of lattice-reduction-aided broadcast precoding \cite{vp_lll} applied to our MSVP method, which uses a matrix decomposition of ${\bf T}$ based on the LLL lattice reduction algorithm \cite{LLL} and similarly computes a Babai estimate of the optimal perturbation vector.  In the bottom plot of Fig.\ \ref{ML_vs_SQR_vs_LLL} we repeat the experiment for $K = 80$ and $c = 1$, $4/5$, omitting the performance of ML MSVP as using a sphere decoder for such a large system is infeasible.

As we see from the plots, the performance degradation of using the SQR method instead of an ML method to find the perturbation vector ${\bf x}$ is minimal.  Surprisingly, the SQR method offers a marginal but consistent improvement over the LLL method at high values of $\rho$.  This is especially notable since, as we discuss further in Section \ref{complexity}, computing the SQR matrix decomposition can be done with lower complexity than computing the LLL reduction.


\subsection{Comparison with Zero-Forcing Vector Perturbation}

In this subsection we compare MSVP (with $\alpha = \sigma^2$) with the zero-forcing strategy in which ${\bf A}= {\bf H}_{\text{ZF}}$ and the offset vector ${\bf x}$ is chosen using the SQR algorithm of Section \ref{sqr} to minimize $\gamma = ||{\bf A({\bf u+x})}||^2/K$ as in \cite{swindlehurst2,heath_se}.  We denote this strategy ZFVP.  Channel pre-inversion with ${\bf A}= {\bf H}_{\text{MMSE}}$ and no perturbation is also shown as a helpful basis for comparison.  

\begin{figure}[h!]
\centering
\includegraphics[width=.40\textwidth]{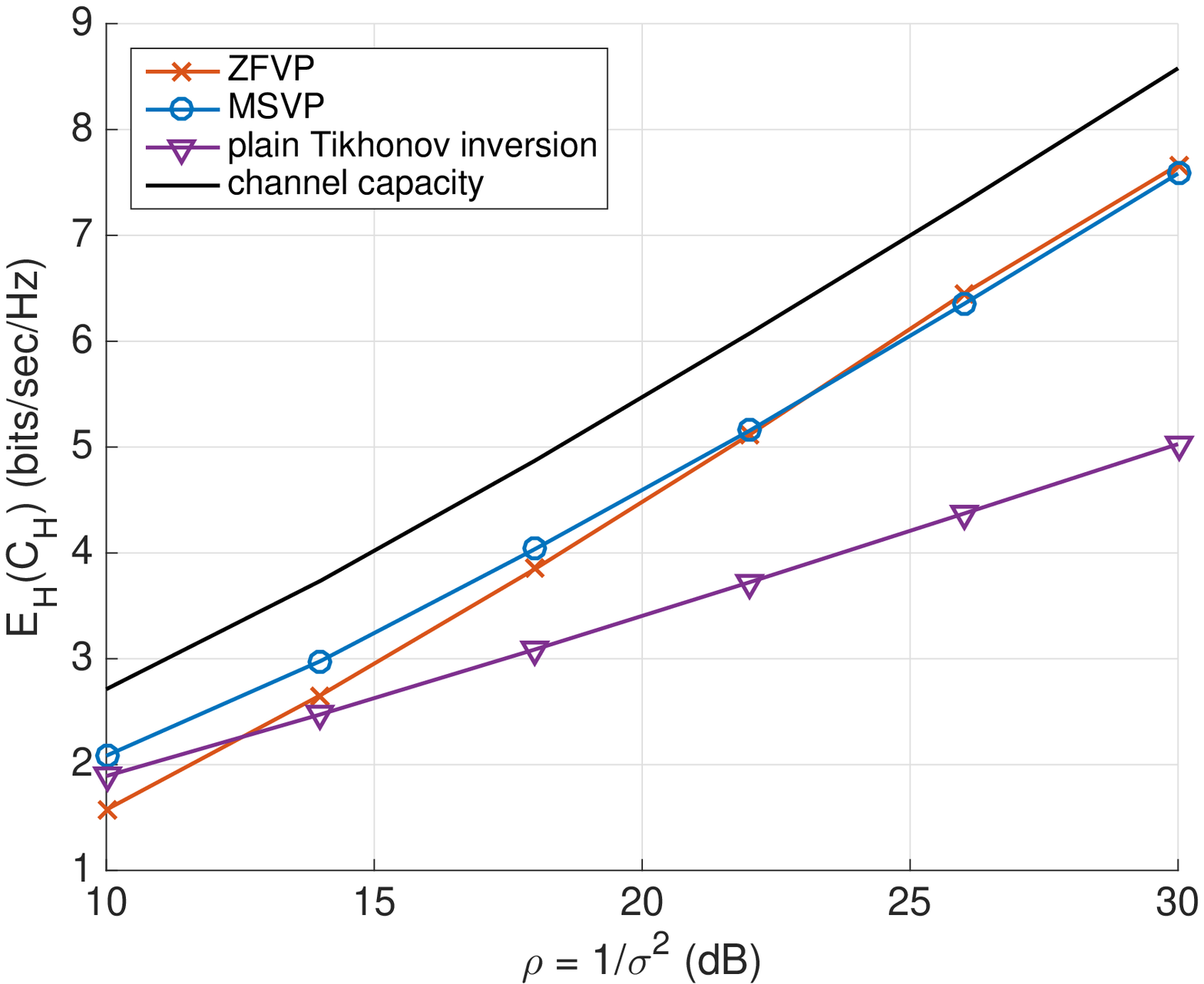} \hfill
\includegraphics[width=.40\textwidth]{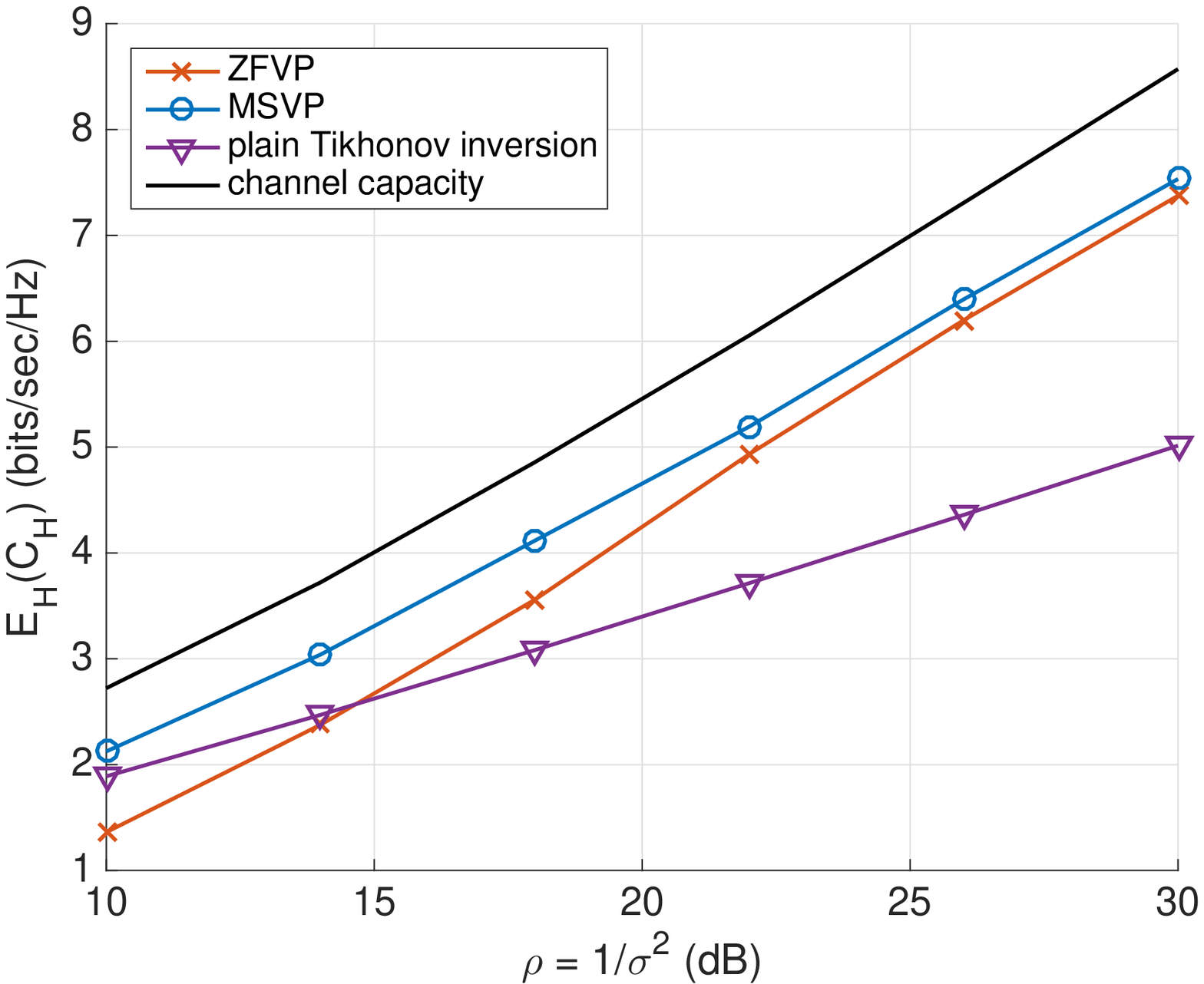}
\caption{On top, the value of $C = \mathbb{E}_{\bf H}(C_{\bf H})$ for the MSVP and ZFVP strategies with $K = M = 256$ employing $16$-QAM signaling, using the SQR decomposition to solve for the perturbation vector.  On bottom, the same plot for $K = M = 1024$.}\label{VP_caps_bigK}
\end{figure}

In Fig.\ \ref{VP_caps_bigK} we show the performance of ZFVP and MSVP for $K = M = 256$ (top) and $K = M = 1024$ (bottom) when the SQR algorithm is employed.  When $K = M = 256$, MSVP offers a steady improvement of approximately $1$ dB over ZFVP between $\rho = 10$ dB and $20$ dB, though for large values of $\rho$ ZFVP slightly outperforms MSVP.  For $K=M=1024$ the performance of ZFVP degrades to the point where we see that MSVP outperforms it at all values of $\rho$ under consideration.  On the contrary, the performance of MSVP is apparently constant with increasing system size.  To simplify presentation we omitted the performance of SQR WFVP from these plots, but the behavior is essentially identical to that already depicted in Fig.\ \ref{VP_caps_K12M12_MSVP_vs_WFVP}.  Specifically, MSVP outperforms WFVP by approximately $0.5$-$1$ dB at all values of $\rho$ when the SQR algorithm is employed to find the perturbation vectors.

\subsection{Remarks on Complexity}\label{complexity}

To demonstrate that SQR MSVP can be used in practice, we now briefly discuss the complexity of the involved algorithms.  Solving for the perturbation vector ${\bf x}$ is the main bottleneck to implementing VP systems, as it must be done multiple times per channel realization and finding the ML solution is notoriously complex.  The preprocessing performed on the channel matrix ${\bf H}$ (e.g.\ computing the matrix ${\bf T}$) must only be done once per channel realization and therefore has less of an impact on total computation.  Nevertheless, we discuss both aspects below.

\subsubsection{Preprocessing} During SQR MSVP, the preprocessing consists of two parts, namely computing the $K\times K$ matrix ${\bf T}$ as in Theorem \ref{Tmatrix}, and then performing the SQR decomposition to ${\bf T}$.  The former can be done with a Cholesky decomposition, the computation of which requires $O(K^3)$ operations.  The latter can be done using a modified Gram-Schmidt algorithm (see \cite{wubben}), the complexity of which is easily be seen to be $O(K^3)$, and therefore all preprocessing can be performed in $O(K^3)$ operations.  If LLL reduction is used instead of the SQR method, then between $O(K^4)$ and $O(K^5)$ operations are needed \cite{LLL}.

\subsubsection{Solving for the Perturbation Vector} Solving for the perturbation vector ${\bf x}$ in (\ref{best_x}) using a sphere decoder has complexity which is exponential in the dimension $K$ of the lattice \cite{sd_complexity}. On the other hand, computing the Babai estimate ${\bf x}^B$ using (\ref{babai1}) and (\ref{babai2}) requires only $O(K^2)$ multiplications.  The complexity of computing the Babai point is the same, regardless of whether we use the SQR or LLL method to compute the perturbation vector.

\section{Conclusions and Future Work}\label{conclusions}

With the goal of developing scalable and close-to-capacity data transmission schemes for next-generation broadcast networks, we have studied channel pre-inversion and vector perturbation schemes for a large number $K$ of end users.  To that end, we have provided an explicit, sharp estimate of the capacity of MMSE channel pre-inversion as $K\rightarrow \infty$.  Furthermore, we have proposed a new max-$\SINR$ vector perturbation scheme which maximizes a sharp estimate of the $\SINR$ of the system.  Random Matrix Theory was used to estimate the performance of our vector perturbation scheme, and the resulting approximation was shown to be accurate.  We demonstrated that MSVP outperforms other VP schemes, such as Wiener Filter VP and zero-forcing VP.  Lastly, we applied the Sorted QR decomposition method to solve for the perturbation vector, resulting in a scheme which is low-complexity and close to channel capacity for very large $K$.  The low complexity and good performance suggest that our max-$\SINR$ vector perturbation method could be implemented in practice in large broadcast networks.


Future work will consist of investigations into using fast-decodable space-time codes \cite{fd_stbc_rajan,fd_stbc_viterbo,fd_stbc_cam} at the transmit end, which could naturally reduce the complexity of the ML search for the perturbation vector.  Furthermore, we plan on comparing our method with the Degree-2 Sparse VP method of \cite{yima}, which offers comparable complexity.  We plan to investigate the performance of MSVP with imperfect CSI and with correlated channel coefficients, particularly when correlation occurs between the transmit antennas.  Lastly, preliminary simulation results suggest that the quantity $\dg({\bf HA})_{ii}/\sqrt{\gamma}$, which must be known by the receivers prior to transmission, is nearly constant with respect to ${\bf H}$, especially for large systems.  Thus it may be possible to replace this quantity with a simple constant at the receive end, cutting down on preliminary communication overhead significantly.  This is an avenue of potential future research that deserves investigation.


%

\appendix

\section{Proof of Theorem 1}\label{proof_theorem1}
\emph{Proof of Theorem 1:}  To prove that the matrix ${\bf T}$ exists, which must first show that the matrix $d^2{\bf I}_K -2d{\bf HH}_\alpha + {\bf HH}_\alpha {\bf H}_{\text{MMSE}}^+{\bf H}_\alpha$ is positive definite.  To that end, let ${\bf H} = {\bf U\Sigma V^\dagger}$ be a singular value decomposition of the channel, where the diagonal entries of ${\bf \Sigma}$ are $s_1,\ldots,s_K$.  The singular value decomposition of ${\bf H}_\alpha$ is
\begin{eqnarray}
{\bf H}_\alpha &=& {\bf V\Sigma}^T {\bf U^\dagger} (\alpha {\bf I}_K + {\bf U\Sigma V^\dagger V}{\bf \Sigma}^T {\bf U}^\dagger)^{-1} \\
&=& {\bf V}\underbrace{{\bf \Sigma}^T(\alpha {\bf I}_K + {\bf \Sigma\Sigma}^T)^{-1}}_{=:{\bf \Sigma}_\alpha}{\bf U}^\dagger \\
&=& {\bf V\Sigma_\alpha U^\dagger}
\end{eqnarray}
where ${\bf \Sigma}_\alpha$ has the singular values $s_i/(s_i^2 + \alpha)$, for $i = 1,\ldots,K$ of ${\bf H}_\alpha$ along the diagonal.  It follows immediately that the singular values of ${\bf H}_{\text{MMSE}}^+$ are $(s_i^2+\sigma^2)/s_i^2$ for $i = 1,\ldots,K$.

Each singular value $s$ of ${\bf H}$ gives rise to an eigenvalue $\lambda$ of the matrix $d^2{\bf I}_K -2d{\bf HH}_\alpha + {\bf HH}_\alpha {\bf H}_{\text{MMSE}}^+{\bf H}_\alpha$, which is given by
\begin{eqnarray}
\lambda &=& d^2 -2d\frac{s^2}{s^2+\alpha} + \frac{s^2(s^2+\sigma^2)}{(s^2+\alpha)^2} \\
&=& d^2 - 2d\frac{s^2}{s^2+\alpha} + \left(\frac{s^2}{s^2+\alpha}\right)^2\frac{s^2+\sigma^2}{s^2} \\
&>& d^2 - 2d\frac{s^2}{s^2+\alpha} + \left(\frac{s^2}{s^2+\alpha}\right)^2  \\
&=& \left(d-\frac{s^2}{s^2+\alpha}\right)^2 \geq 0.
\end{eqnarray}
As all $\lambda$ are obtained this way, the matrix $d^2{\bf I}_K -2d{\bf HH}_\alpha + {\bf HH}_\alpha {\bf H}_{\text{MMSE}}^+{\bf H}_\alpha$ is positive definite.

Let $\varepsilon_1$ and $\varepsilon_2$ be as in the statement of the theorem.  To see the bounds for the $\SINR$, note that when ${\bf A} = {\bf H}_\alpha$ we have
\begin{align}
&\SINR \nonumber \\
&= \frac{||\dg({\bf HH}_\alpha)||^2_Fc}{\mathbb{E}_{\bf u}\left(||({\bf HH}_\alpha-\dg({\bf HH}_\alpha))({\bf u+x})||^2+ ||{\bf H_\alpha({\bf u+x})}||^2\sigma^2\right)} \\
&= \frac{||\dg({\bf HH}_\alpha)||^2_Fc}{\mathbb{E}_{\bf u}\left(||({\bf HH}_\alpha-d{\bf I}_K)({\bf u+x})||^2+ ||{\bf H_\alpha({\bf u+x})}||^2\sigma^2\right)+\varepsilon_2} 
\end{align}
We begin by bounding the numerator above and below.    By the triangle inequality, we have
\begin{eqnarray}
||\dg({\bf HH}_\alpha)||^2_F &=& ||\dg({\bf HH}_\alpha) - d{\bf I}_K + d{\bf I}_K||^2_F\\
&\leq& ||d{\bf I}_K||^2_F + \varepsilon_1 \\
&=& d^2K + \varepsilon_1
\end{eqnarray}
To see the other inequality, let $a_1,\ldots,a_K>0$ be any positive real numbers, and let $a = \frac{1}{K}\sum_{i = 1}^K a_i$ be their mean.  We have $(\sum_{i = 1}^Ka_i)^2\leq (\sum_{i = 1}^Ka_i^2)(\sum_{i = 1}^K1^2)$ by the Cauchy-Schwarz Inequality, which is easily seen to be equivalent to $||[a,\ldots,a]^T||^2\leq ||[a_1,\ldots,a_K]^T||^2$.  Letting $a_i = ({\bf HH}_\alpha)_{ii}$ and $a = d = \tr({\bf HH}_\alpha)/K$ we see that $d^2K = ||d{\bf I}_K||^2_F\leq ||\dg({\bf HH}_\alpha)||^2_F$.

To complete the bounds on the $\SINR$, define
\begin{equation}
\widehat{\MSE}_{\bf u} = ||({\bf HH}_\alpha-d{\bf I}_K)({\bf u+x})||^2+ ||{\bf H_\alpha({\bf u+x})}||^2\sigma^2
\end{equation}
Following \cite[Section 4]{wiener_se}, the idea is to rewrite $\widehat{\MSE}_{\bf u}$ as the norm of a single vector.  To shorten notation, we let ${\bf z = u+x}$ and ${\bf z' = U^\dagger z}$.  Computing $\widehat{\MSE}_{\bf u}$ in terms of the singular value decompositions now gives (noting that multiplying by a unitary matrix does not affect the norm of a vector)
\begin{align}
\widehat{\MSE}_{\bf u} 
&= ||({\bf \Sigma\Sigma}_\alpha-d{\bf I}_K){\bf z}'||^2 + ||{\bf \Sigma}_\alpha {\bf z}'||^2 \\
&= \sum_{i = 1}^K\left(\frac{s_i^2}{s_i^2+\alpha} - d\right)^2|z_i'|^2 + \sum_{i = 1}^K\left(\frac{s_i}{s_i^2+\alpha}\right)^2|z_i'|^2\sigma^2 \\
&= \sum_{i = 1}^K\left(d^2 -2d\frac{s_i^2}{s_i^2+\alpha} + \frac{s_i^2(s_i^2+\sigma^2)}{(s_i^2+\alpha)^2}\right)|z_i'|^2
\end{align}
Now suppose that ${\bf T}$ is as in the statement of the proposition.  Using the singular value decompositions of ${\bf H}$ and ${\bf H}_\alpha$, we see that an eigenvalue decomposition of ${\bf T}^\dagger {\bf T}$ is given by
\begin{align}\label{eig_decomp}
{\bf T}^\dagger {\bf T} &= d^2{\bf I}_K -2d{\bf HH}_\alpha + {\bf HH}_\alpha {\bf H}_{\text{MMSE}}^+{\bf H}_\alpha  \\
&= {\bf U}\ \text{diag}\left(d^2 -2d\frac{s_i^2}{s_i^2+\alpha} + \frac{s_i^2(s_i^2+\sigma^2)}{(s_i^2+\alpha)^2}\right){\bf U}^\dagger
\end{align}
where for any $a_i \in \mathbb{C}, i = 1,\ldots,K$ we define $\text{diag}(a_i)$ to be the diagonal matrix with the vector $[a_1,\ldots,a_K]$ along the diagonal.  Comparing this computation with the previous one shows that
\begin{align}
||{\bf T}({\bf u+x})||^2 &= {\bf z}^\dagger {\bf T}^\dagger {\bf T} {\bf z}   \\
&= ({\bf z}')^\dagger \text{diag}\left(d^2 -2d\frac{s_i^2}{s_i^2+\alpha} + \frac{s_i^2(s_i^2+\sigma^2)}{(s_i^2+\alpha)^2}\right){\bf z}'  \\
&= \widehat{\MSE}_{\bf u}
\end{align}
which completes the proof of the bounds on the $\SINR$. 

It remains to prove that $\lim\limits_{K\rightarrow\infty}\frac{1}{K}\mathbb{E}_{\bf H}(\varepsilon_1) = 0$.  Let ${\bf U} = (u_{ij})$ be as in the singular value decomposition of ${\bf H}$.  An elementary matrix computation gives
\begin{align}
\varepsilon_1 &= \sum_{i = 1}^K|\dg({\bf HH}_\alpha)_{ii} - d|^2 \\
&= \sum_{i = 1}^K\left| \sum_{j = 1}^K\left(|u_{ij}|^2-\frac{1}{K}\right)\frac{s_j^2}{s_j^2+\alpha}\right|^2 \\
&\leq \sum_{i = 1}^K\left(\sum_{j = 1}^K\left||u_{ij}|^2-\frac{1}{K}\right|^2\right)\left(\sum_{j = 1}^K\left(\frac{s_j^2}{s_j^2+\alpha}\right)^2\right) \\
&\leq \sum_{i = 1}^K\sum_{j = 1}^K\left||u_{ij}|^2-\frac{1}{K}\right|^2 \leq \sum_{i = 1}^K\sum_{j = 1}^K|u_{ij}|^4 + 3.
\end{align}
The first inequality follows from the Cauchy-Schwarz Inequality applied to the previous line, the second inequality is obvious, and the third inequality follows from the triangle inequality and the unitarity of ${\bf U}$.  From this we see that it suffices to show $\lim\limits_{K\rightarrow\infty}\frac{1}{K}\mathbb{E}_{\bf H}\left(\sum_{i,j}|u_{ij}|^4\right) = 0$.

Let $\mathbb{U}(K)$ be the group of complex unitary $K\times K$ matrices, which is compact and therefore admits a uniform distribution, which comes from the Haar measure.  Consider the distribution on $\mathbb{U}(K)$ obtained from the random matrix ${\bf U}$ coming from the singular value decomposition ${\bf H} = {\bf USV}$ of the random Gaussian matrix ${\bf H}$. If ${\bf Q}$ is any other unitary matrix, then ${\bf QH} = {\bf QUSV}$ has the same distribution as ${\bf H}$, hence ${\bf QU}$ has the same distribution as ${\bf U}$.  Since the Haar measure is the unique translation-invariant measure on $\mathbb{U}(K)$, it follows that ${\bf U}$ defines the uniform distribution on $\mathbb{U}(K)$.  For ${\bf U}$ uniform on $\mathbb{U}(K)$, we have by \cite[Lemma 2.5]{verdu_rmt} that $\mathbb{E}|u_{ij}|^4 = \frac{2}{K(K+1)}$. It follows immediately that $\lim\limits_{K\rightarrow\infty}\mathbb{E}_{\bf H}\left(\sum_{i,j} |u_{ij}|^4\right) = 2$, which concludes the proof of the theorem.  \hfill$\blacksquare$

\section{Proof of Lemma 2}{\label{proof_lemma1}
\emph{Proof of Lemma 2:}  Let us write $E_K = \mathbb{E}(X_K)$ to shorten notation.  We rewrite $\frac{X_K}{1-X_K}$ as
\begin{align}
\frac{X_K}{1-X_K} &= \frac{X_K}{1-E_K}\frac{1}{1-\frac{X_K-E_K}{1-E_K}} \\
&= \frac{X_K}{1-E_K}\sum_{k = 0}^{\infty}\left(\frac{X_K-E_K}{1-E_K}\right)^k
\end{align}
where the validity of the geometric series expansion follows from assumption (i).  Taking the expectation of the above and simplifying the first few terms gives
\begin{equation}\label{exp_power_series}
\begin{aligned}
\mathbb{E}\left(\frac{X_K}{1-X_K}\right) &= \frac{E_K}{1-E_K} + \frac{\text{Var}(X)}{(1-E_K)^2} \\
&+ \sum_{k = 2}^{\infty}\frac{\mu_{k+1}(X_K) + E_K\mu_k(X_K)}{(1-E_K)^3}
\end{aligned}
\end{equation}
where $\mu_k(X) = \mathbb{E}(X-\mathbb{E}(X))^k$, so that $\mu_2(X) = \text{Var}(X)$.  By Jensen's Inequality and assumption (i) we have $|\mu_k(X_K)| \leq \mathbb{E}|(X_K-E_K)^k| \leq \text{Var}(X_K)$.  By assumption (iii) we have $\lim\limits_{K\rightarrow \infty}\mu_k(X_K) = 0$ for all $k\geq 2$.  Taking the limit as $K\rightarrow \infty$ of (\ref{exp_power_series}) and using assumption (ii) to guarantee this is well-defined gives the result. \hfill $\blacksquare$

\section{Proof of Theorem 2}\label{proof_theorem2}
\emph{Proof of Theorem 2:}  Let us first compute $\lim\limits_{K\rightarrow\infty}\mathbb{E}_{\bf H}(d)$, which is a straightforward application of a result from Random Matrix Theory.  Suppose that $f:[0,\infty)\rightarrow \mathbb{C}$ is bounded and continuous, and let $\lambda_i$ be the $i^{th}$ eigenvalue of $ {\bf HH}^\dagger$.  Then by \cite[Corollary 7.8]{haagthor}, we have
\begin{equation}\label{rmt_result1}
\lim_{K\rightarrow \infty}\frac{1}{K}\sum_{i = 1}^Kf(\lambda_i) = \frac{1}{c}\int_a^bf(x/c)\frac{\sqrt{(x-a)(b-x)}}{2\pi x}dx
\end{equation}
where $a = (\sqrt{c}-1)^2$ and $b = (\sqrt{c}+1)^2$.  To apply this result, we write $d$ as
\begin{align}
d &=  \frac{1}{K}\tr({\bf HH}^\dagger(\alpha {\bf I}_K+{\bf HH}^\dagger)^{-1}) \\
&= \frac{1}{K}\sum_{i = 1}^K\frac{\lambda_i}{\lambda_i+\alpha}
\end{align}
and thus
\begin{equation}
\lim_{K\rightarrow\infty}\mathbb{E}_{\bf H}(d) = \lim_{K\rightarrow\infty}\frac{1}{K}\sum_{i = 1}^Kf(\lambda_i),
\end{equation}
where $f(x) = \frac{x}{x+\alpha}$.  Plugging the function $f(x)$ into (\ref{rmt_result1}) completes the proof of this part of the theorem, as Mathematica readily evaluates the corresponding integral to be equal to $d(c,\alpha)$.

We note that it is easy to see $0<d<1$ for all $K$, and additionally a simple computation gives $0<d(c,\alpha)<1$.  The result thus follows immediately from Lemma 1, provided that $\lim\limits_{K\rightarrow\infty}\text{Var}(d) = 0$.  But $d$ is the mean of the empirical eigenvalue distribution of ${\bf H}{\bf H}_\alpha$ \cite[Section 1.2]{verdu_rmt}, and hence as $K\rightarrow \infty$ the pdf of $d$ converges almost surely to the Dirac delta distribution centered at the mean of the Marshenko-Pasteur distribution \cite[(1.10)]{verdu_rmt}.  In other words, the limiting distribution of $d$ as $K\rightarrow \infty$ is a point mass distribution, which has zero variance.  It is easy to verify that $\lim\limits_{K\rightarrow\infty}\text{Var}(d)$ is equal to the variance of this limiting distribution, which completes the proof.

The lower bound on $\widehat{C}$ follows from a simple application of Jensen's Inequality.

\bibliographystyle{ieee}
\bibliography{myrefs_new}

\end{document}